\documentclass[aps,prb,twocolumn,superscriptaddress,longbibliography]{revtex4-2}

\usepackage[utf8]{inputenc}

\usepackage{physics}
\usepackage{comment}
\usepackage{amssymb}

\usepackage[colorlinks=true,citecolor=blue]{hyperref} 
\usepackage{graphicx}
\usepackage{amsmath}
\usepackage{dcolumn}
\usepackage{bm}

\newcommand{\newc}{\newcommand}
\newc{\kt}{\rangle}
\newc{\br}{\langle}
\newc{\pr}{\prime}
\newc{\longra}{\longrightarrow}
\newc{\ot}{\otimes}
\newc{\rarrow}{\rightarrow}
\newc{\h}{\hat}
\newc{\bom}{\boldmath}
\newc{\btd}{\bigtriangledown}
\newc{\al}{\alpha}
\newc{\be}{\beta}
\newc{\ld}{\lambda}
\newc{\sg}{\sigma}
\newc{\p}{\psi}
\newc{\eps}{\epsilon}
\newc{\om}{\omega}
\newc{\mb}{\mbox}
\newc{\tm}{\times}
\newc{\hu}{\hat{u}}
\newc{\hv}{\hat{v}}
\newc{\ii}{\dot{\iota}}
\newc{\cf}{{\cal{F}}}
\newc{\md}{\mbox{D}}
\newc{\RNum}[1]{\uppercase\expandafter{\romannumeral #1\relax}}

\begin{document}

\title{Non-Hermitian topology and criticality in photonic arrays with engineered losses} 

\author{Elizabeth Louis Pereira}

\affiliation{Department of Applied Physics, Aalto University, 02150 Espoo, Finland}

\author{Hongwei Li}
\affiliation{Nokia Bell Labs, 21 JJ Thomson Avenue, Cambridge, CB3 0FA, UK}

\author{Andrea Blanco-Redondo}
\affiliation{CREOL, The College of Optics and Photonics, University of Central Florida, Orlando, FL 32816, USA}

\author{Jose L. Lado}
\affiliation{Department of Applied Physics, Aalto University, 02150 Espoo, Finland}

\date{\today}

\begin{abstract}
Integrated photonic systems provide a flexible platform where artificial lattices can be engineered in a reconfigurable fashion. Here, we show that one-dimensional photonic arrays with engineered losses allow the realization of topological excitations stemming from non-Hermiticity and bulk mode criticality. We show that a generalized modulation of the local photonic losses allows the creation of topological modes both in the presence of periodicity and even in the quasiperiodic regime. We demonstrate that a localization transition of all the bulk photonic modes can be engineered in the presence of a quasiperiodic loss modulation, and we further demonstrate that such a transition can be created in the presence of both resonance frequency modulation and loss modulation. We finally address the robustness of this phenomenology to the presence of next to the nearest neighbor couplings and disorder in the emergence of criticality and topological modes. Our results put forward a strategy to engineer topology and criticality solely from engineered losses in a photonic system, establishing a potential platform to study the impact of nonlinearities in topological and critical photonic matter.
\end{abstract}

\maketitle

\section{Introduction}

Topological insulators are one of the emerging platforms to study novel phenomena in quantum matter\cite{PhysRevLett.61.2015,RevModPhys.83.1057,RevModPhys.82.3045}. 
Topological modes have been realized in a variety of artificial systems including mechanical\cite{Huber2016}, photonic\cite{Wang2009,Khanikaev2012}, and cold atom setups\cite{Jotzu2014}.
Topological photonics\cite{RevModPhys.91.015006,price2022} has risen as a powerful platform to generate
new states of light that harvest non-trivial
geometric properties in lasers\cite{Bandres2018,Contractor2022} and quantum information platforms\cite{BlancoRedondo2018,Mittal2018,Wang2019}.
Topological states can emerge in systems lacking a periodic
lattice, including disordered models\cite{PhysRevResearch.2.013053,Pyhnen2018,Marsal2020} and quasicrystals\cite{PhysRevLett.53.1951,PhysRevLett.53.2477,aubry,PhysRevLett.115.195303,PhysRevLett.109.106402,PhysRevResearch.2.022049,PhysRevB.98.125431,PhysRevB.91.064201,Zilberberg2021}, featuring criticality stemming from localization transitions\cite{Goblot2020}.
Photonic devices allow the creation of a whole variety
of new artificial lattices\cite{PhysRevB.91.064201,Goblot2020,Zilberberg2018} challenging to emulate in conventional
materials, opening up possibilities to realize new forms of topological matter.

\begin{figure}[t!]
\includegraphics[width=\linewidth]{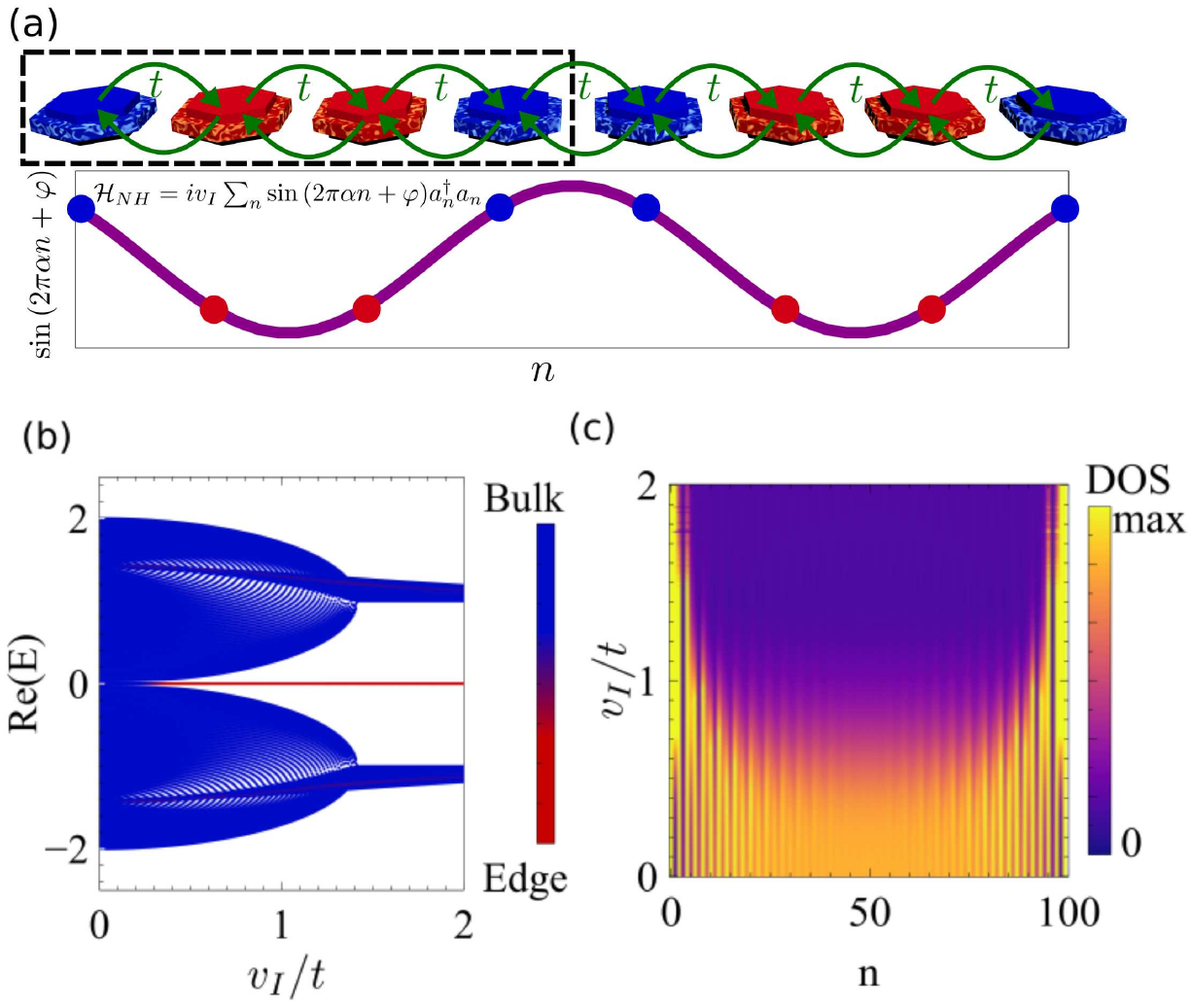}
\caption{(a) Schematic of the gain-loss model with a unit cell of four sites (dashed rectangle). 
The local loss is modulated by the function $\ii v_{n}=\ii v_{I}\sin\left(2\pi \alpha n+ \varphi\right )$, with $\alpha=1/4$ and $\varphi =3\pi /4$. 
(b) Real part of the energy spectrum as a function of $v_{I}/t$ for the model shown in (a), for $N=200$. The spectrum is symmetric with respect to $\mbox{Re}(E)=0$ with an edge mode present at $\mbox{Re}(E)=0$, which is topologically robust for the entire range of $v_{I}/t$. (c)The density (DOS) of the zero mode is shown as a function of $v_{I}/t$ and the site index $n$. With an increase in the value of $v_{I}$, the zero mode gets exponentially localized at the end sites. We took $N=100$ for (c).  }
\label{panel1}
\end{figure}

Beyond conventional photonic
topological states in closed quantum systems\cite{Khanikaev2012,StJean2017},
photonic devices provide a flexible platform to harvest non-Hermitian topology\cite{Okuma2023,RevModPhys.93.015005,PhysRevLett.121.026808,PhysRevB.84.205128,PhysRevLett.80.5243,Denner2021},
and in particular, robust topological modes by 
exploiting engineered gains and losses\cite{PhysRevResearch.4.L012006,takata,ghatak,PhysRevLett.109.106402}.
Integrated reconfigurable\cite{Zhao2019,Shalaev2018,Arora2021,Bogaerts2020,PrezLpez2020,Harris2018,Clements2016,andreas} photonic devices provide 
a flexible platform to engineer tunable photonic matter by allowing real-time reconfiguration of optical paths.
This tunability turns reconfigurable photonic devices 
into an ideal platform to explore exotic topological phenomena in a non-Hermitian and spatially engineered
regimes\cite{PhysRevB.104.024201,PhysRevB.103.014203,PhysRevB.100.125157}.

In this manuscript, we present a strategy to engineer
topological modes and criticality simultaneously
in one-dimensional photonic arrays solely based
on engineered losses. 
In particular, we show that a generalized set of models with engineered losses feature topological edge modes
stemming from non-Hermitian topology. For quasiperiodic modulations, 
we show that the modulated losses lead to a delocalization to localization transition
of the bulk states.
We analyze the resilience of the topological edge modes to disorder in the engineered losses and detuning frequency,
and the impact of long-range tunneling in the localization transition and topological modes.
Our results provide a starting point for designing topological
photonic devices based on tunable losses. 
Our manuscript is organized as follows. In section \ref{sec:model}, we present the generalized model
featuring modes from engineered losses. In section \ref{sec:crit}, we analyze the localization transition
driven by modulated losses. In Sec. \ref{sec:dis}, we address the impact of perturbations and disorder.
In Sec. \ref{sec:cont} we address the continuum limit of the model. 
Finally, in Sec. \ref{sec:con}, we summarize our conclusions.

\section{Topological modes from engineered losses}
\label{sec:model}
We consider a one-dimensional array of photonic dots featuring localized excitation. Photonic losses are included
by adding a non-Hermitian term into a 1D model in each site of the array. 
For the sake of concreteness, we first consider an engineered loss with four-site periodicity\cite{takata,PhysRevB.100.161105,PhysRevB.101.180303,PhysRevApplied.13.014047,PhysRevResearch.4.L012006,PhysRevLett.130.100401}, as shown in Fig.[\ref{panel1}(a)], whose Hamiltonian takes the form  
\begin{equation}
H = t\sum\limits _{n=0}^{N-2}\left( a_{n}^{\dagger}a_{n+1} + h.c. \right ) + i\sum\limits_{n=0}^{N-1}
(v_0 + v_{I}\beta_n )a_{n}^{\dagger}a_{n},
\label{ham1}
\end{equation}

where
$a^\dagger_n$ creates photon in site $n$, $v_0 + v_{I}\beta_n$, with $v_0, v_I $
real numbers denotes the site-dependent loss, parametrised by the modulation ($\beta_1 = \beta_4 = 1$) and ($\beta_2 = \beta_3 = -1$). The term $v_0$ leads to an overall loss in the system, and therefore in the following it will
be factored out in the spectra. Due to the non-Hermiticity of the Hamiltonian, the eigenenergies $E_\alpha$ will be in general complex, with
$H |\Psi_\alpha \rangle = E_\alpha |\Psi_\alpha \rangle$.
We show in Fig.[\ref{panel1}(b)] the real part of the energy spectrum of this model for open boundary conditions as a function of $v_{I}$. We can see that the edge modes with $\mbox{Re}(E)=0$ are strongly localized at the edges, and represent the topological edge modes arising from the modulated
loss. The extent of localization for the edge modes is shown in Fig.[\ref{panel1}(c)] using the
spectral function at zero real energy $D_0(n)=\sum_\alpha \delta(\mbox{Re}(E_{\alpha})) | \Psi_\alpha (n) |^2$.
In particular, as loss modulation strength increases the edge modes get localized 
at the end sites as compared to those in the bulk.

\begin{figure}[t!]
\includegraphics[width=\linewidth]{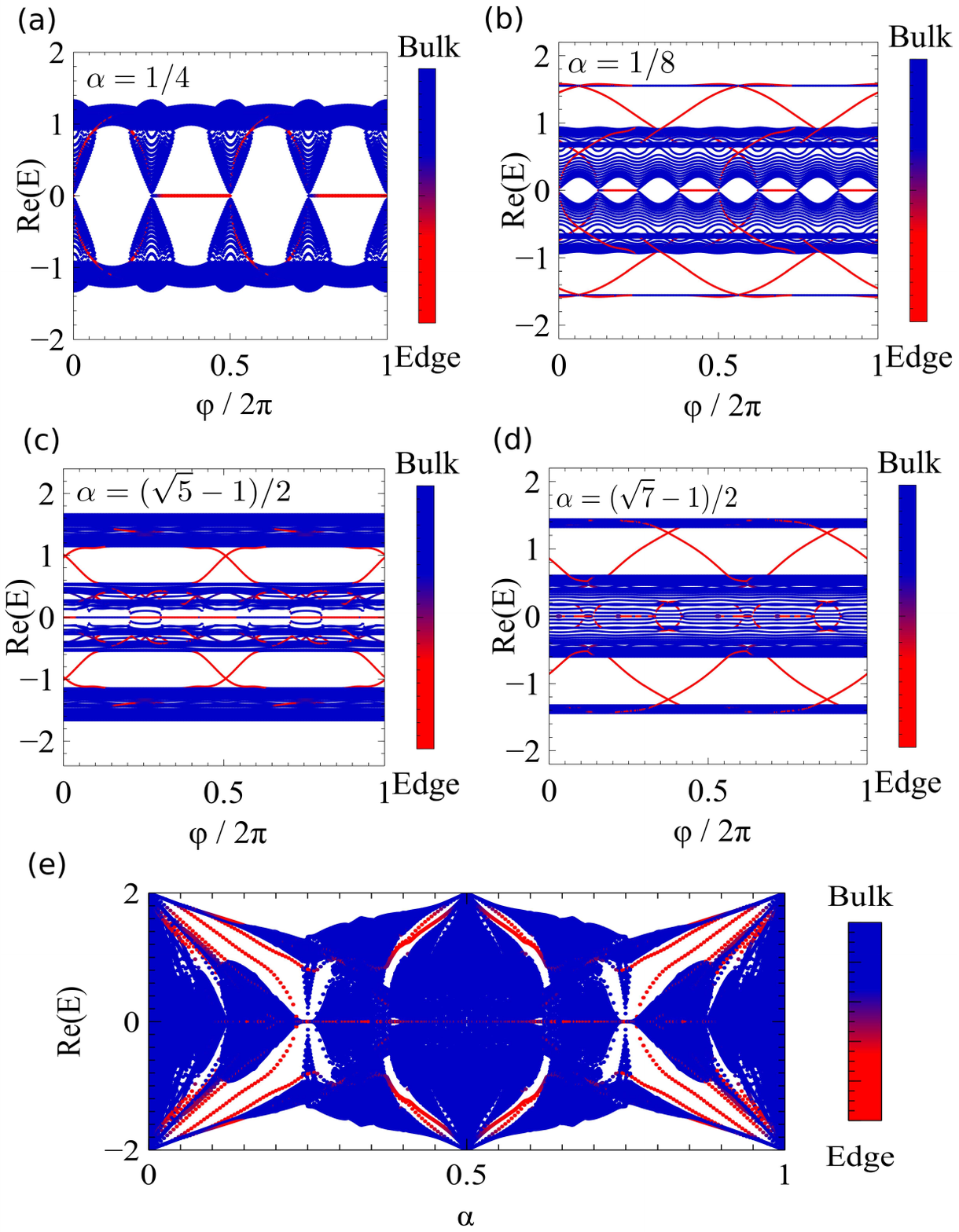}
\caption{ 
(a-d) Spectra as a function of $\varphi$ for a chain with modulated losses, for $\alpha=1/4$ (a), $\alpha=1/8$ (b),
and in the incommensurate limit $\alpha=(\sqrt{5}-1)/2$ (c), and $\alpha=(\sqrt{7}-1)/2$ (d).
The energies are colored according to the spatial location of the state in the chain.
It is observed that
edge modes appear in spectral gaps for wide regions of $\varphi$. 
Panel (e) shows the spectra as a function of the wavevector of the modulation of the loss $\alpha$, showing that
topological edge modes appear for generic values inside the spectral gap.
We took $v_{I}=1.5$ and $N=200$ for (a-e).}

\label{panel2}
\label{fig:fig2}
\end{figure}

The previous topological non-Hermitian model given by Eq.[\ref{ham1}] can be seen as a specific case of a generalized non-Hermitian model given by the following Hamiltonian
\begin{equation}
H = t \sum\limits_{n=0}^{N-2}\left( a_{n}^{\dagger}a_{n+1} + h.c.\right) 
+ i \sum\limits_{n=0}^{N-1} v_{I}\sin(2\pi n \alpha + \varphi) a_{n}^{\dagger}a_{n},
\label{ham2}
\end{equation}
where $\alpha$ is the inverse period of modulation, $\varphi$ is the phase of modulation, and $v_{I}$ is the amplitude of modulation of the losses. 
This model can be realized as the non-Hermitian generalization of the Aubry-Andr\'{e}-Harper (AAH) potential, \cite{aubry}.
In its Hermitian form $H = t \sum\limits_{n=0}^{N-2}\left( a_{n}^{\dagger}a_{n+1} + h.c.\right) 
+ \sum\limits_{n=0}^{N-1} v\sin(2\pi n \alpha + \varphi) a_{n}^{\dagger}a_{n}$, this model is well known to be equivalent
to a two-dimensional quantum Hall system, \cite{PhysRevLett.109.106402}, thus inheriting topological edge modes. However, such a mapping cannot be performed in its
non-Hermitian generalization. As it is shown in Fig.[\ref{panel1}(a)], the systems in Eq.[\ref{ham1}] and in Eq.[\ref{ham2}] are equivalent when  $\alpha = \frac{1}{4}$ and $\varphi=\frac{3\pi}{4}$. 
To have a periodic system, the potential should be commensurate to the lattice periodicity.
In particular, when $\alpha$ is a rational number $p/q$, the total number of sites, $N$ should be a multiple of $q$. 
While for a quasiperiodic system, $\alpha$ should be an irrational number. 
For different values of $\alpha$ and $\varphi$, we obtain a series of topologically inequivalent insulators \cite{zhulang}.

We now show how topological modes can appear for the generic values of the parameters $\alpha$ and $\varphi$\cite{Ganeshan2013,zhulang}.
We show in Fig.[\ref{panel2}(a-d)], the real part of the energy spectrum for different values of the parameter $\alpha$. 
In the case of $\alpha=1/4$ (Fig.[\ref{panel2}(a)]), the topological edge modes emerge at zero energy. In contrast, topological modes at other
values of $\alpha$ appear at finite energies.
Also, the real part of the bulk modes is symmetric with respect to the band gap
at $\mbox{Re}(E)=0$ which is a feature of particle-hole symmetry. The
particle-hole symmetry is associated with the system when $q$ is a multiple of
$4$. Figs.[\ref{panel2}(a),(b)] show the commensurate limit\cite{zhulang},
where the frequency of the modulation leads to a periodic system. 
Figs.[\ref{panel2}(c),(d)] are for incommensurate frequencies, we can see that the energy spectra have a fractal nature for these cases and have edge modes shown in red.
Also, many band gaps in the energy spectrum do not have robust edge states, as
can be inferred from the figure. Note, all these systems shown in
Fig.[\ref{panel2}(a-d)] an have imaginary component of the energy that is not
shown in the figure.

We can also study the presence of edge modes for systems with a range in $\alpha$ by computing the spectra of the system as a function of the modulation frequency $\alpha$. In its Hermitian version, such a plot is known as the Hofstadter butterfly spectrum of the Hamiltonian. 
We show the real part of the Hofstadter spectrum of Hamiltonian $H$ given by Eq.[\ref{ham2}] in Fig.[\ref{panel2}(e)]. 
The edge modes are plotted in red which appear in band gaps for various $\alpha$, showing the appearance of those modes even for modulation
frequencies not commensurate with the lattice.

\section{Criticality and localization-delocalization transition}
\label{sec:crit}

\begin{figure}[t!]
\includegraphics[width=\linewidth]{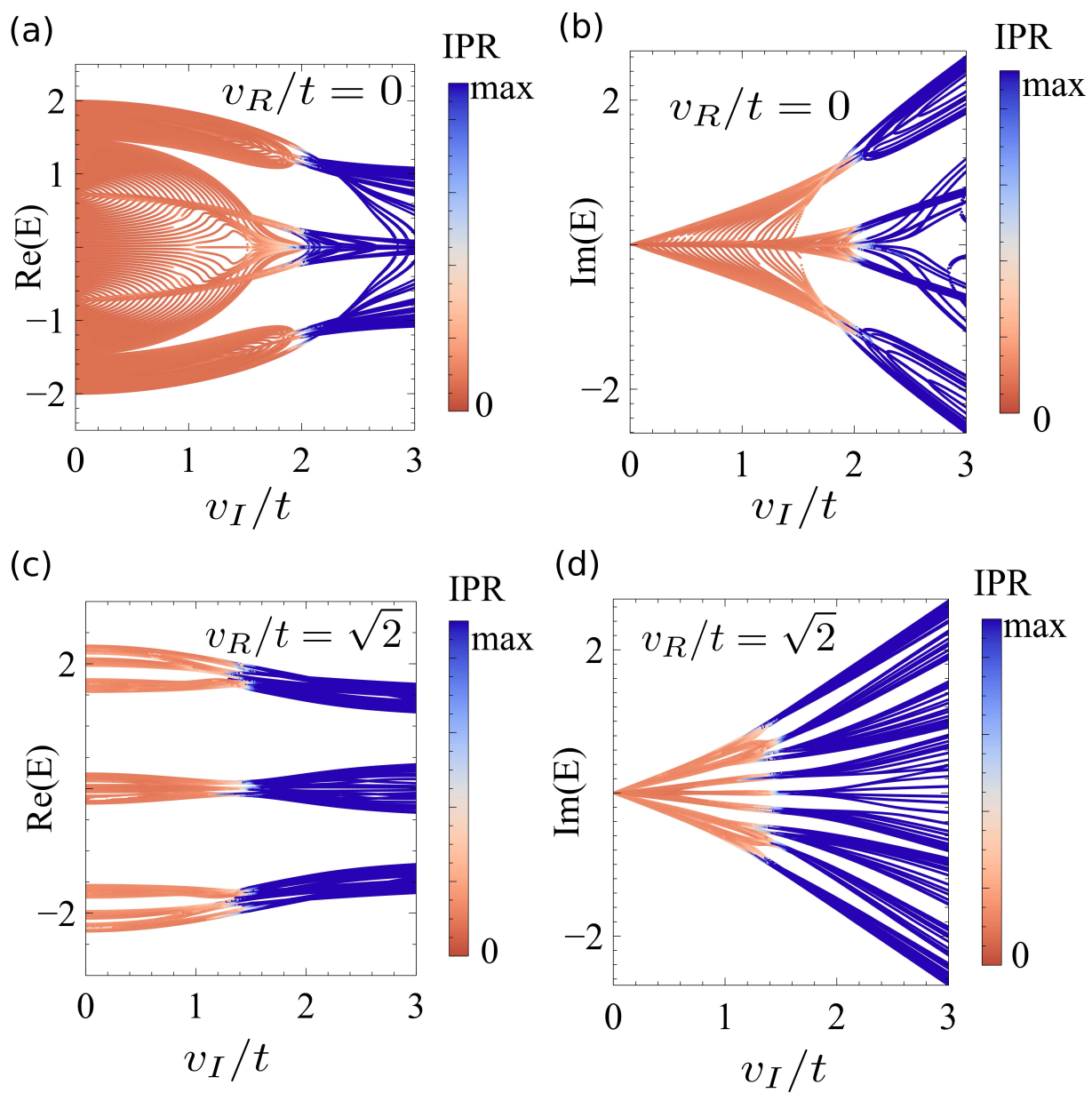}
\caption{(a-d) Spectra for a chain model as a function of onsite losses $v_{I}$, with IPR as the color. 
Panels (a,b) show that the localization transition occurs at $v_{I}=2t$ when $v_{R} =0$, whereas panels (c,d)
show that the localization transition occurs at $v_{I}=\sqrt{2}t$ when $v_{R} =\sqrt{2}t$. 
We took $N=200$, $\alpha = (\sqrt{5}-1)/2$, and $\varphi = 0.4\pi$ in (a-d).}
\label{panel3}
\label{fig:loc}
\end{figure}

Quasiperiodic Hermitian models feature localization transitions at finite strength, phenomena
that turned them into an attractive platform to realize wavefunction criticality\cite{Jitomirskaya1999,PhysRevLett.125.196604,PhysRevLett.50.1870,PhysRevResearch.3.013262,PhysRevLett.50.1873,Goblot2020}.
The Hermitian AAH model described by an onsite potential $v\sin(2\pi\alpha n +\varphi)$ is known to have a localization transition as a function of the modulation strength $v\in \mathbb{R}$. This model is self-dual and has a limit of self-duality at $v=2t$, i.e. all the bulk states localize at $v=2t$\cite{aubry}. 
The localization transition as a function of $v$ is independent of the phase of modulation $\varphi \in\mathbb{R}$ for a quasiperiodic system\cite{aubry}.
In the following, we study the localization transition of the non-Hermitian AAH model. The localization transition
can be directly inferred from the calculation of the inverse participation ratio (IPR) of the eigenstates\cite{PhysRevB.100.125157}. For a state $|\psi\kt $, the IPR is defined as
\begin{equation}
\mbox{IPR }(|\psi\kt) = \sum\limits _{l} |\psi_{l}|^{4}.
\end{equation}
For $N \rightarrow \infty$, we have $\mbox{IPR } = 0$ for an extended state and
$\mbox{IPR } \sim 1/W$, with $W$ the number of sites where the state is localized, for a localized state.
We study this transition for the system described by Eq.[\ref{ham2}] with respect to the modulation strength $v_{I}$. 
We show in Fig.[\ref{panel3}(a)] and Fig.[\ref{panel3}(b)] the IPR for all the eigenvalues of the Hamiltonian given by Eq.[\ref{ham2}] with respect to the amplitude of modulation $v_{I}$ for $\alpha = (\sqrt{5}-1)/2$. 
We can see that a localization transition for all the eigenstates occurs at $v_{I}=2t$, simultaneously for all eigenstates. As a reference, the maximum value of IPR in the figures is 
of the order $18/N$.
In the Hermitian version of this model, a similar phenomenology takes place stemming
from self-duality between the coordinate and the momentum space. 
It should be contrasted that in the case of conventional disorder,
the localization transition occurs at infinitesimally small disorder
for a one-dimensional model\cite{PhysRev.109.1492,RevModPhys.57.287}. The existence of a critical value directly reflects the inherent quasiperiodicity of the potential, making this model genuinely different from a disordered system.

\begin{figure}[t!]
\includegraphics[width=\linewidth]{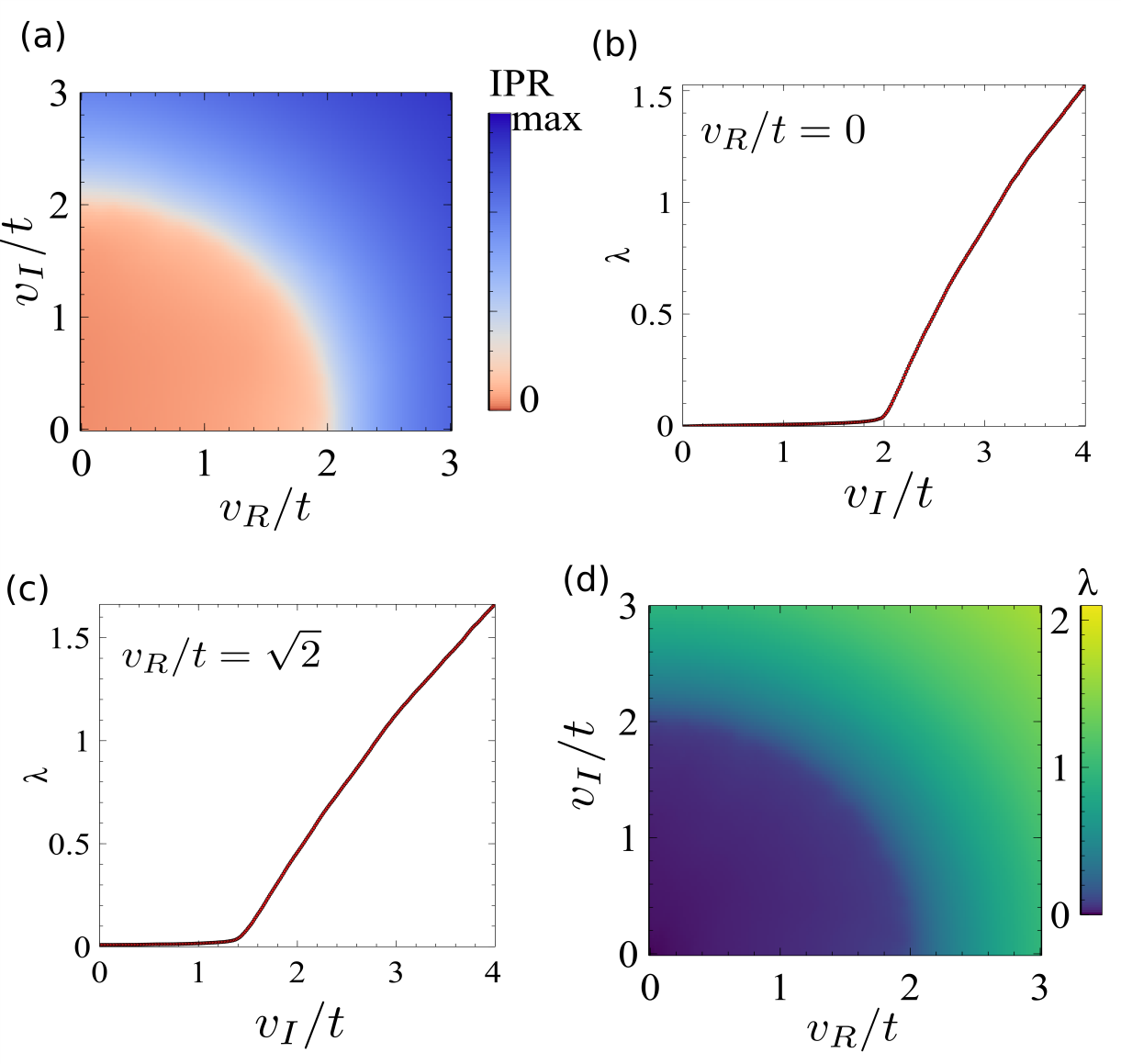}
\caption{(a) The average IPR of the chain model as a function of $v_{R}$ and $v_{I}$, the system gets localized along the contour $v_{R}^{2}+v_{I}^{2}=4t^{2}$. (b,c) The inverse of localization length $\lambda$ as a function of $v_{I}$ for $v_{R}=0$ and $v_{R}=\sqrt{2}t$ respectively showing the critical localization transition at $v_{I} =2t$ and $v_{I}=\sqrt{2}t$ respectively. (d) The inverse of localization length ($\lambda$) of the chain model as a function of $v_{R}$ and $v_{I}$, the system gets localized along the contour $v_{R} ^{2}+v_{I}^{2}=4t^{2}$. We took $\alpha=(\sqrt{5}-1)/2, N=200, \varphi = 0.4 \pi$ in (a-d).}
\label{panel4}
\end{figure}

The existence of a localization transition in a non-Hermitian model, analogous to the one known in its Hermitian counterpart,
motivates the question of whether there exists a generalized model featuring such a localization-delocalization.
For this purpose, we now address
the localization transition for a complex modulation strength. Consider the following modification for our Hamiltonian,
\begin{eqnarray}
H &=& t \sum\limits_{n=0}^{N-2}\left( a_{n}^{\dagger}a_{n+1} + h.c.\right) \nonumber \\
&& +  \sum\limits_{n=0}^{N-1}(v_{R}+\ii v_{I})\sin(2\pi n \alpha + \varphi) a_{n}^{\dagger}a_{n},
\label{ham4}
\end{eqnarray}
where $v_{I}$ is the modulation of the loss and $v_{R}$ the modulation of the onsite potential of the system, such that $v_{R}, v_{I}$ are real parameters.
In the case $v_I = 0$, the previous model is equivalent to the Hermitian AAH model,
whereas for $v_R = 0$ we recover our model with modulated losses.
We start by fixing $v_{R}$ as a nonzero value to study the spectrum as a function of the loss modulation $v_{I}$.  
We show in Fig.[\ref{panel3}(c)] and Fig.[\ref{panel3}(d)] the energy as a function of $v_{I}$ for $v_{R}=\sqrt{2} t$. We observe that approximately at $v_{I}=\sqrt{2} t$, all the eigenstates undergo a localization-delocalization transition.
A finite value of the onsite quasiperiodicity $v_{R}$, leads to 
a different critical value for localization transition as a function of the quasiperiodic engineered loss $v_{I}$.
To elucidate how the critical transition depends on both modulations, 
we show in Fig.[\ref{panel4}(a)] a two-dimensional phase diagram
as given by the IPR as a function of both $v_{R}$ and $v_{I}$ for $\alpha = (\sqrt{5}-1)/2$.  
It is observed that a localization-delocalization transition occurs following the approximate critical line
$v_{R}^2 + v_{I}^2 = 4t^2$. As a reference, the critical transition in the Hermitian AAH model corresponds to the cut $v_{I}=0$, whereas
the critical transition in the purely imaginary model corresponds to the cut in $v_{R}=0$. This phenomenology
highlights that 
the Hermitian AAH model belongs to a general family of non-Hermitian AAH models with complex modulation strength. 

The localization-delocalization transition can also be studied from the localization length of the wavefunction. In the localized limit,
localized eigenstates can be fitted to a functional form such as 
\begin{equation}
  |\psi _{\alpha}|^{2} (n) \sim  e^{-\lambda |r_{0}-n|},
  \label{funcform}
\end{equation}
where $\alpha$ labels the eigenstate, $n$ is the site of the chain
and $r_{0}$ is the center of eigenstate i.e. where $ |\psi _{\alpha}|^{2} (n) $ is the highest. The parameter
$1/\lambda$ is the localization length, which in the case of an extended state corresponds to $1/\lambda=\infty$. For each eigenstate of the system, we perform a fit to the previous functional form (Eq.[\ref{funcform}]), which allows extracting a localization length for each state.
We show in Fig.[\ref{panel4}(b)] the average $\lambda$ versus $v_{I}$ for $v_{R}=0$, showing that a localization-delocalization transition
occurs at $v_{I}=2t$. Similarly, setting $v_{R}$ as $\sqrt{2}t$, we see that the localization occurs at $v_{I}=\sqrt{2}t$ as in Fig.[\ref{panel4}(c)]. In Fig.[\ref{panel4}(d)], we show a phase diagram according to the inverse localization length as a function of the parameters $v_{R}$ and $v_{I}$, 
where we can see that the localization occurs at $v_{R}^{2}+v_{I}^{2}= 4t^{2}$. The localization length is given as a function of the modulation strength, by the Thouless formula as $\frac{1}{\lambda}=\frac{1}{\log\left(|v|/(2t)\right)}$, where $v=v_{R}+i v_{I}$ and $|v|>2$, \cite{aubry}.

\section{Long-range coupling and disorder}
\label{sec:dis}
\subsection{Impact of long-range coupling}

So far our analysis has focused
on the limit featuring first nearest-neighbor coupling. In realistic
experimental scenarios, finite coupling between longer neighbors may occur. 
Couplings beyond first neighbors are expected to give rise to an energy-dependent
localization transition. In particular, the Hermitian AAH
model with an exponentially decreasing hopping $t e^{-p|n -n'|}a^\dagger_n a_{n'}$
gives rise to an energy-dependent localization
transition that can be derived analytically\cite{Biddle2010}.
The non-Hermitian limit however cannot be addressed with
the self-duality procedure of the
Hermitian limit, and thus we will focus here on an exact numerical strategy.
For this purpose, we now address the impact of second and third-nearest-neighbor coupling. 
The Hamiltonian for this system takes the form

\begin{eqnarray}
    H  &=& t\sum_{n=0}^{N-2}\left(a_{n}^{\dagger}a_{n+1}+a_{n+1}^{\dagger}a_{n}\right)
    \nonumber \\ &&+ \sum_{n=0}^{N-1}\left(v_{R}+\ii v_{I}\right)\sin\left(2\pi \alpha n+\varphi\right)a_{n}^{\dagger}a_{n}\nonumber \\ &&+t_{2}\sum_{n=0}^{N-3}\left(a_{n}^{\dagger}a_{n+2}+a_{n+2}^{\dagger}a_{n}\right)\nonumber \\ &&+t_{3}\sum_{n=0}^{N-4}\left(a_{n}^{\dagger}a_{n+3}+a_{n+3}^{\dagger}a_{n}\right).
    \label{hamihigherneighbhopping}
\end{eqnarray}

\begin{figure}[t!]
\includegraphics[width=\linewidth]{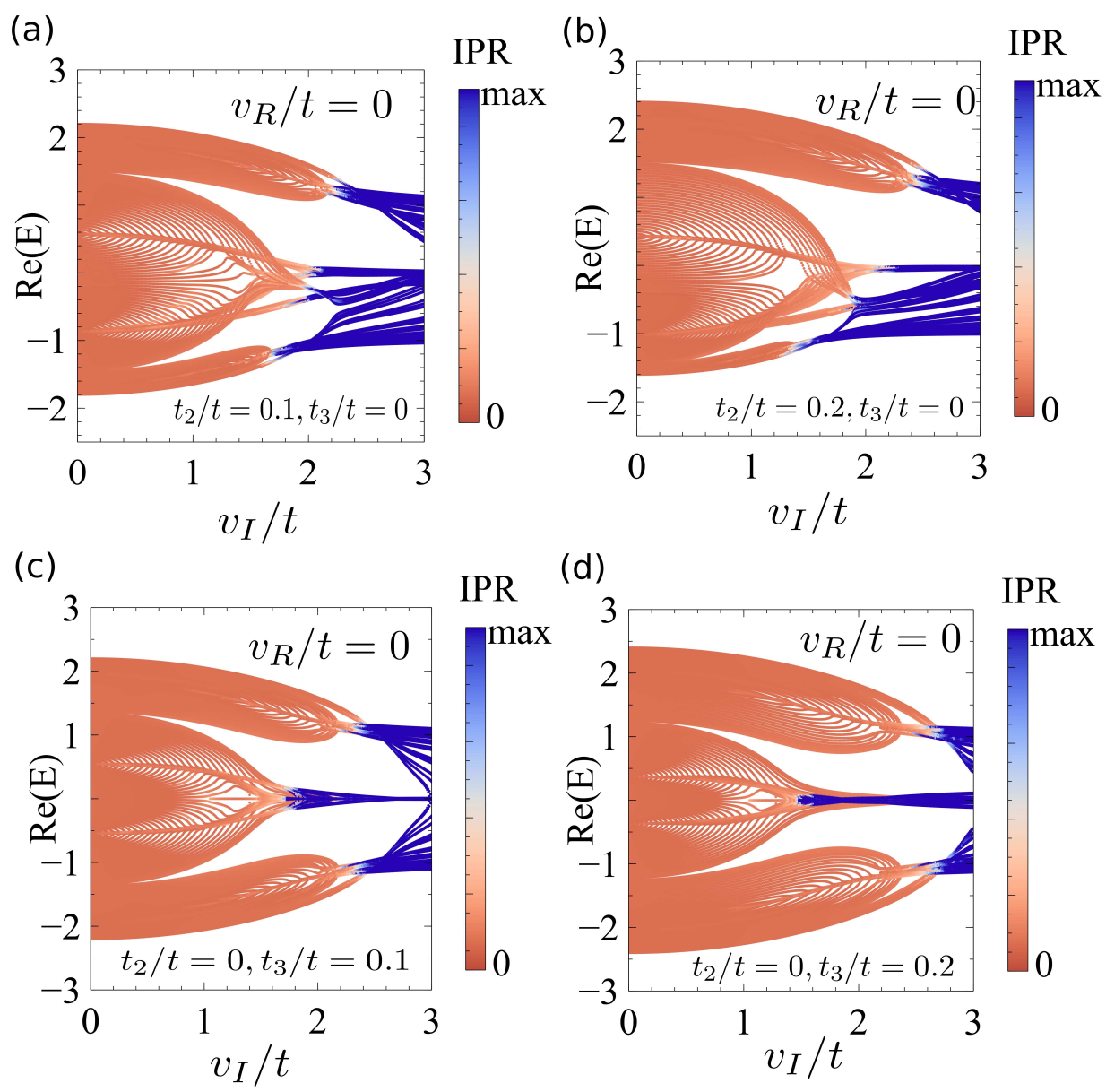}
\caption{(a-d) Real part of energy spectra for the chain model with losses 
as a function of the strength of modulation $v_{I}$
for different values of the second and third neighbor hoppings
$t_2$ and $t_3$. The localization transition occurs for different eigenstates at different $v_{I}$. 
The inclusion of finite $t_2$ shows that the spectra have no particle-hole symmetry  (a,b), 
whereas third neighbor hopping preserves particle-hole symmetry (c,d). 
We took $N=200,v_{R}=0,\alpha =(\sqrt{5}-1)/2$ and $\varphi=0.4\pi$ in (a-d).}
\label{fig:nnn}
\end{figure}

We show in Fig.[\ref{fig:nnn}], the real part of the energy spectrum for the Hamiltonian given by Eq.[\ref{hamihigherneighbhopping}] with
respect to the modulation strength $v_{I}$ when $v_{R}=0$. 
Panels 
Fig.[\ref{fig:nnn}(a)] and Fig.[\ref{fig:nnn}(b)] 
show the evolution of the localization of the bulk modes as a function of $v_{I}$ in the presence of second nearest neighbor hopping $t_{2}$.
It is observed that the inclusion of second neighbor hopping removes the particle-hole symmetry of the spectra. 
Interestingly, we observe that at certain energies such as $v_I=1.9t$, localized and extended states coexist in the bulk.
This must be contrasted with the situation observed in Fig.[\ref{fig:loc}] where it was observed that all the eigenstates are either
extended or localized, and no coexistence is possible. The coexistence of localized and extended states is associated
with a mobility edge, and in Fig.[\ref{fig:nnn}(a,b)] we observe that this mobility edge depends on the strength $t_{2}$,
appearing in a wider region for increasing $t_{2}$. 
It is instructive to address another case with extended hopping, in particular, third neighbor hopping $t_3$
as shown in Fig.[\ref{fig:nnn}(c,d)], while taking $t_2=0$.
In this scenario, we observe that the spectrum remains particle-hole symmetric to $\mbox{Re}(E)=0$. 
This extended model also features a localization transition as a function of $v_{I}$, happening at different parameter values depending on the 
energy. 
In particular, the bulk states with energy $|\mbox{Re}(E)|$ closer to $0$ get localized for a smaller $v_{I}$ as compared to
those that are farther. 
Similar phenomenology is observed in a fully Hermitian model, where mobility edge exists for the second and third neighbour hopping.
These results highlight that higher-order couplings lead to an energy-dependent localization of the non-Hermitian
bulk states, regardless of whether
they maintain the particle-hole symmetry of the underlying model.

\begin{figure}[t!]
\includegraphics[width=\linewidth]{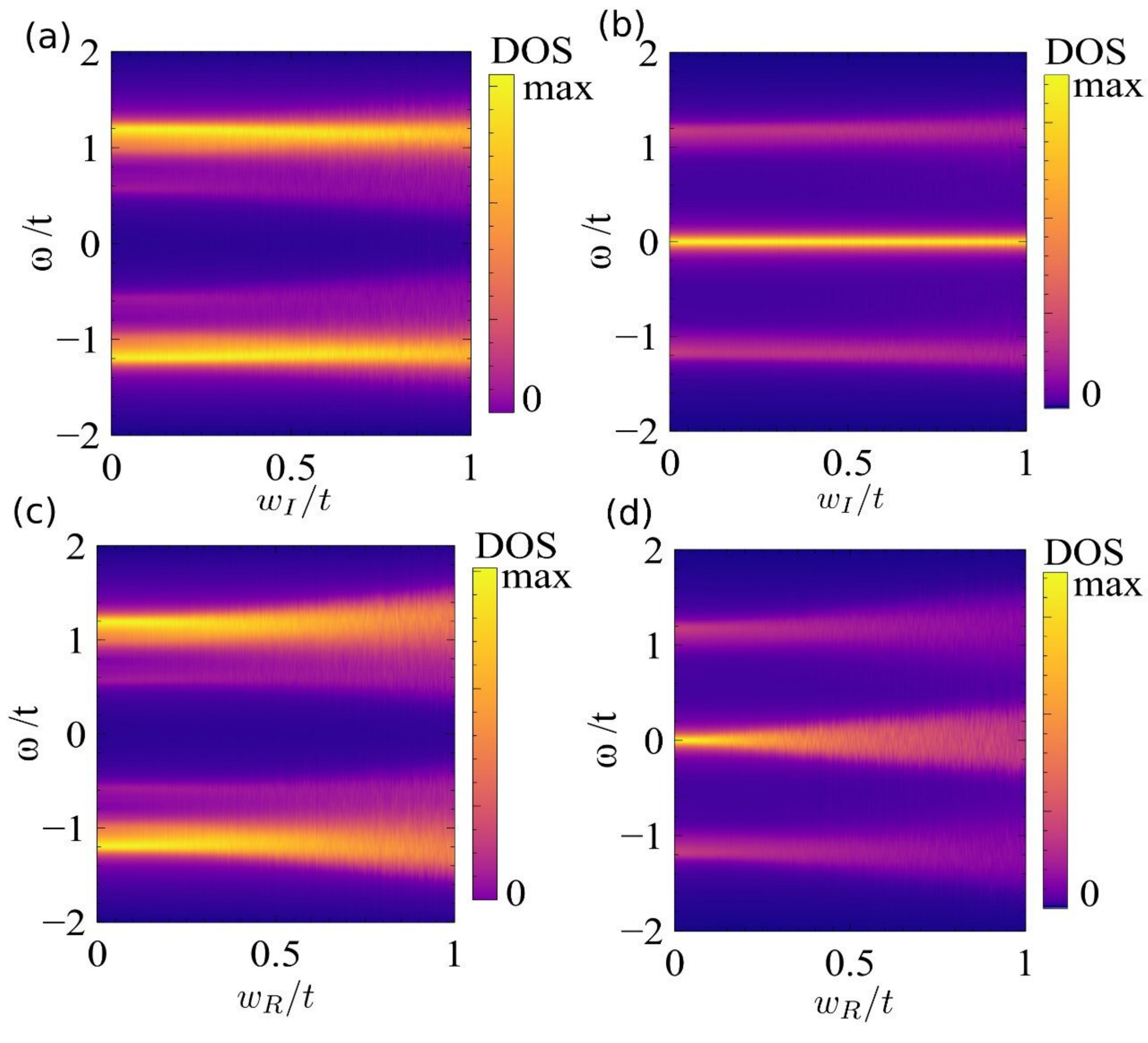}
\caption{(a-d) Spectral density as a function of the loss disorder strength $w_I$ (a,b) and potential disorder $w_R$ (c,d), in the bulk (a,c)
and at the edge (b,d). It is observed that the existence of a finite disorder decreases the bulk gap (a,c), but without destroying it.
The zero edge modes are robust to the existence of disorder in loss as shown (b), whereas they develop a finite splitting
in the presence of detuning disorder (d).
We took $N=200$ and $v_{I}=2$, and results are averaged over $100$ realizations.}
\label{panel6}
\end{figure} 

\subsection{Impact of disorder}

We now study the robustness of the system given by Eq.[\ref{ham2}] as a function of disorder in
the imaginary onsite energy \cite{zhulang}.
For the sake of concreteness, we will focus on the model featuring modulation solely on the losses by
taking $v_{R}=0$ and $v_{I}$ is nonzero corresponding to the model
featuring modes at zero energy for $\alpha=1/4$. 
To study the effect of disorder on the energy spectrum, 
we define the spatially resolved spectral density as
\begin{equation}
D(\omega,n)=\sum_\alpha \delta(\omega - \mbox{Re}(E_{\alpha})) | \Psi_\alpha (n) |^2 ,
\end{equation}
where $\omega$ is the frequency. We note that the previous quantity projects onto the real part of the eigenenergy, whereas an analogous one can be defined for the imaginary part. The spectral density $D(\omega)$ provides direct access to the number of eigenstates with a specific value
in the real part of the energy, and is analogous to the density of states in a Hermitian system.

We study the spectral density for the Hamiltonian given by Eq.[\ref{ham2}] with respect to the disorder strength,
considering loss disorder and detuning disorder.
The disorder is included in the Hamiltonian
by including a term
\begin{equation}
H_D = w_R
\sum_n \chi_{n,R} a^\dagger_n a_n
+
i w_I
\sum_n \chi_{n,I} a^\dagger_n a_n ,
\end{equation}
where $w_{I}$ and $w_{R}$ parametrize the loss and detuning disorder, respectively.
The disorder is included by sampling a Gaussian distribution $\chi$ with an average value 0
and width $1$, and for the sake of simplicity we will consider
loss and detuning disorder separately.
Let us first focus on the disorder in the loss modulation. 
We show in Fig.[\ref{panel6}(a,b)] 
the spectral function averaged over disorder in the loss modulation
projected on the bulk (Fig.[\ref{panel6}(a)]) and at the edge (Fig.[\ref{panel6}(b)]).
It is observed that the spectral gap in the bulk remains open in the presence of disorder (Fig.[\ref{panel6}(a)]),
and that a robust zero mode remains at the edge even at finite disorder (Fig.[\ref{panel6}(b)]).
This phenomenology highlights that the topological zero mode is robust to the presence of disorder in the loss.
We now move on to the situation where disorder appears in the resonance frequency of each site (Fig.[\ref{panel6}(c,d)]).
We show in Fig.[\ref{panel6}(c,d)] 
the spectral function averaged over disorder in the detuning disorder
projected on the bulk (Fig.[\ref{panel6}(c)]) and at the edge (Fig.[\ref{panel6}(d)]).
It is observed that this disorder also keeps the spectral gap open in the bulk as shown in (Fig.[\ref{panel6}(c)]).
However, it is observed that at the edge the energy of the zero modes is no longer pinned at zero energy, leading to an edge
state at a finite energy proportional to the typical level of the disorder ((Fig.[\ref{panel6}(d)])).
This phenomenology highlights that while disorder in the loss does not impact the resonance frequency of the edge mode,
disorder in the onsite energies affects the topological edge mode.

\section{The Aubry-Andr\'{e}-Harper model in the continuum limit}
\label{sec:cont}

Previously we have focused on the discrete AAH model. In the following, we now bring our attention to the non-Hermitian AAH model in the continuum limit. The continuous AAH model is given by the Hamiltonian in continuous space as
\begin{equation}
\mathcal{H} = 
\int \left [ \frac{\hat p^2}{2m} + (v_{R}+i v_{I})\cos(2\pi \alpha x) \right ] \Psi^\dagger_x \Psi_x d x
\end{equation}
where $\hat p = -i \frac{\partial}{\partial_x}$, $m$ the effective mass, $\alpha$ is the AAH frequency
and $\Psi^\dagger_x$, $\Psi_x$ continuum field operators fulfilling $[ \Psi_x , \Psi^\dagger_{x'} ] = \delta(x-x')$. 
The eigenbasis of the previous Hamiltonian can be computed by solving the associated Sturm–Liouville non-Hermitian differential
equation of the form $\left [ -\frac{\partial^2_x}{2m} + (v_{R}+i v_{I})\cos(2\pi \alpha x) \right ] \psi_k (x) = \epsilon_k \psi_k (x)$,
with $\epsilon_k$ the complex eigenvalue and $k$ parametrizing the phase picked due to twisted boundary
conditions $\psi_k(x + 1/\alpha) = e^{i k}\psi_k(x)$.
It is worth noting that in this continuum limit, the absence of an underlying lattice makes the Hamiltonian
explicitly periodic in space, with a periodicity $1/\alpha$.

\begin{figure}[t!]
\includegraphics[width=\linewidth]{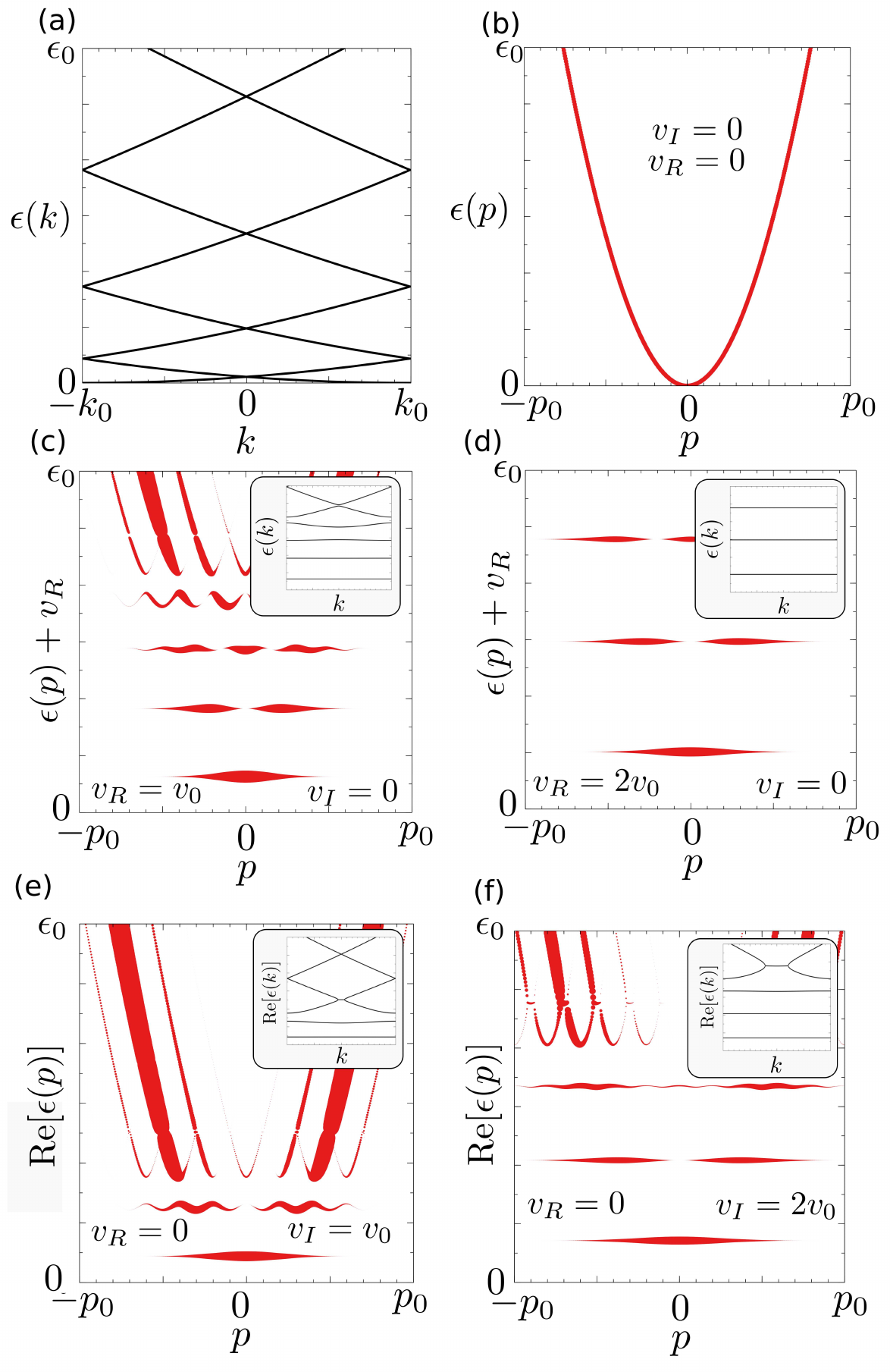}
\caption{
(a) Electronic dispersion in the folded momentum space $k$ for $v_R=v_I = 0$, and
(b) unfolded dispersion in the momentum space $p$ recover the free gas dispersion $p^2/(2m)$.
Panels  (c,d) show the spectra in the unfolded momentum $p$ for the continuum Hermitian AAH model
for two different potential strengths,
showing the appearance of weakly dispersive modes and nearly free states.
The insets in (c,d) show the dispersion in the folded momentum space $k$. 
Panels (e,f) show the spectra for the continuum non-Hermitian AAH model for two potential strengths.
It observed that both weakly dispersive and nearly
free states emerge,
with the inset showing the spectra in the unfolded $k$ space.
}
\label{fig:cont}
\label{panel7}
\end{figure} 

In the absence of a potential $v_R = v_I = 0$, the photonic dispersion $\epsilon (k)$ corresponds to the
folded band structure of a free particle $p^2/(2m)$ as shown in Fig. \ref{fig:cont}(a).
Due to the periodicity of the potential $1/\alpha$, the quasimomentum $k$ must be unfolded to the original free momentum $p$
to recover a parabolic dispersion even for $v_R = v_I = 0$.
For the sake of comparison with the free particle limit, in the following, we will perform an unfolding of the eigenvalues
as a function of the momentum $k$ in the unit cell of size $1/\alpha$, 
which in the case $v_R = v_I = 0$ gives rise to the original band dispersion $\epsilon(p)$ of
Fig. \ref{fig:cont}(b) for the unfolded momentum $p$.
With the previous methodology to solve the continuum model and unfold its eigenvalues
we first address the Hermitian AAH model, and later move to the non-Hermitian version.
Focusing first on the Hermitian case $v_R \ne 0$ and $v_I = 0$, we show in Fig.[\ref{fig:cont}(c,d)]
the unfolded eigenvalues for two strengths of the AAH potential.
The insets of Fig.[\ref{fig:cont}(c,d)] show the photonic dispersion in the original
quasimomentum space $k$ before the unfolding is performed.
It is observed that at the lowest energies, weakly dispersive states appear, giving rise to a set of minibands
with weak dispersion, that eventually lead to a highly dispersive state at high energies
recovering the free particle dispersion. The emergence of those minibands is easily rationalized from the fact that, at
strong $v_R$, the Hamiltonian describes a set of deep harmonic potentials, each one leading to harmonic oscillator modes.
Due to the finite depth of the potential, harmonic oscillator modes between different potential wells have a finite overlap,
leading to a weak dispersion of the individual modes. At higher energies, equivalent to higher modes of the oscillator, the
tunneling between different wells becomes stronger. Finally, at kinetic energies bigger than the depth of the well,
the eigenstates resemble the free particle dispersion.
We now move on to the non-Hermitian AAH continuous model with $v_R = 0$ and $v_I \ne 0$, as shown in
Fig.[\ref{fig:cont}(e,f)]. In this limit, it is observed that at low energies a set of weakly dispersing modes
appear leading to a set of nearly flat mini-bands. At higher energies, the free particle dispersion is recovered, analogously
to the Hermitian case. It is further observed that at high energies a small replica of the dispersion is obtained due to the momentum scattering
created by the non-Hermitian potential.
In contrast with the Hermitian case, the emergence of weakly dispersive states is no longer associated with harmonic oscillator modes in each well,
but they stem purely from the non-Hermitian potential.

\begin{figure}[t!]
\includegraphics[width=\linewidth]{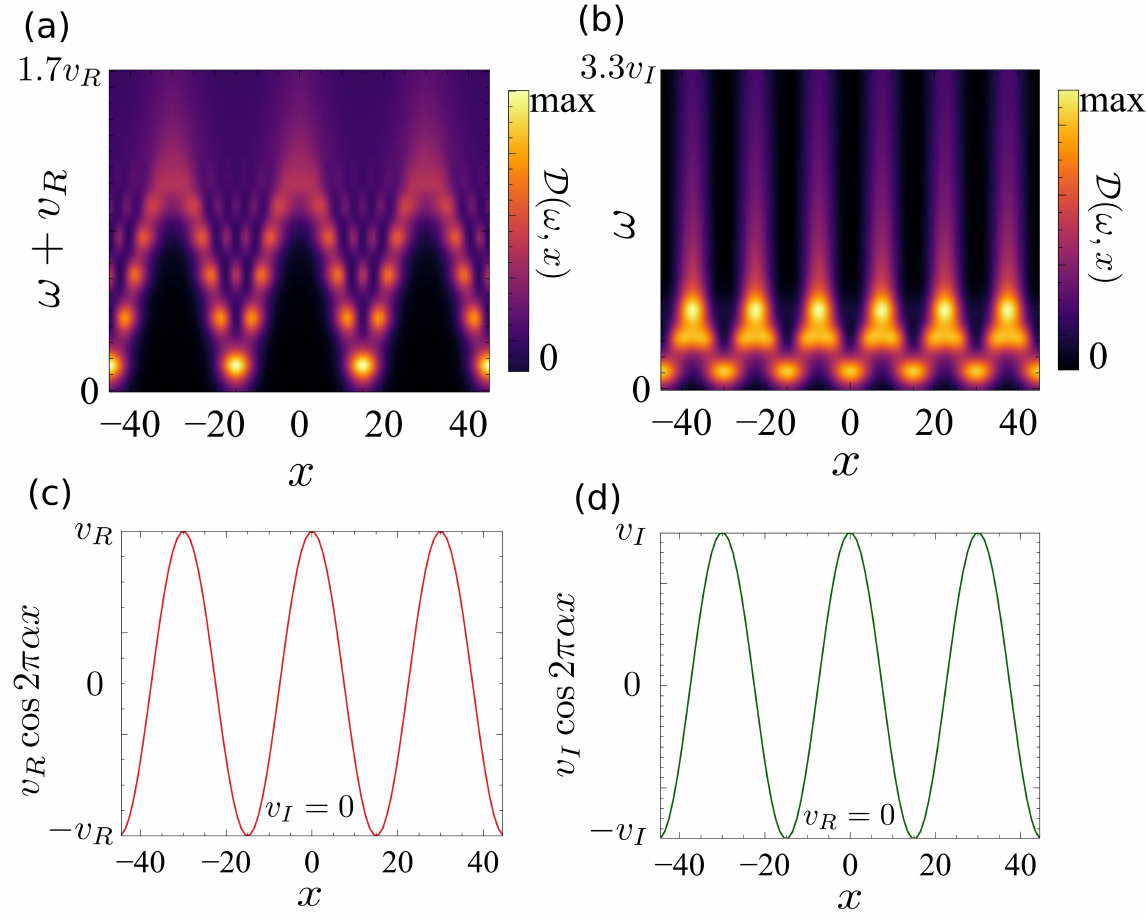}
\caption{(a,b) Spectral density of the Hermitian AAH and non-Hermitian AAH model respectively is shown as a function of space $x$. (c,d) show the potential profile $V(x)$ as a function of $x$ for $v_{I}=0$ and $v_{R}$ nonzero and vice versa. (a) The lower energy states get localized corresponding to $x$ when the potential is zero in (c), with an increase in the value of $\mbox{Re}(E)$ the states are more extended. (b) The lower energy states get localized corresponding to $x$ when the potential attains a $\pm v_{max}$ in (d). Also, with an increase in the value of $\mbox{Re}(E)$, the states are more extended and at a lower value of $\mbox{Re}(E)$ in comparison to the Hermitian case.}
\label{fig:contldos}
\end{figure} 

The nature of the states in Fig.[\ref{fig:cont}] can be further elucidated by computing the spatially resolved continuum
spectral density that takes the form
\begin{equation}
\mathcal{D}(\omega,x)= \int \delta \left [ \omega - \mbox{Re}(\epsilon_{k}) \right ] | \psi_k (x) |^2 dk 
\end{equation}

The spatially resolved spectral density for the continuum Hermitian and non-Hermitian AAH models are shown in Fig.[\ref{fig:contldos}(a,b)],
with Fig.[\ref{fig:contldos}(c,d)] showing the profiles of the Hermitian and non-Hermitian potential as a reference. 
In the Hermitian case shown in Fig.[\ref{fig:contldos}(a) ($v_{I}=0$)], 
it is observed that the low energy states are localized at the bottom of the Hermitian potential wells, as expected from
harmonic oscillator modes.
As the energy is increased, the spatial extension of the modes becomes bigger, leading to an increasing
overlap between the states in the different wells, and accounting for
the enhancement of the bandwidth shown in Fig.[\ref{fig:cont}(c,d)].
At high energies, bigger than the top of the potential, nearly free wavefunctions are recovered extending
through all $x$. We now move on to consider the modes of the non-Hermitian
potential ($v_R = 0$), shown in Fig.[\ref{fig:contldos}(b)]. In the non-Hermitian case, it is observed that the lowest energy modes
are located both at the maxima and minima of the non-Hermitian potential.
At higher frequencies, the extension of the confined modes becomes bigger, leading to the increased bandwidth observed in
Fig.[\ref{fig:cont}(e,f)].
Finally, at high enough energies, the nearly free particle dispersion is recovered, with the unique phenomenology that the states
remain peaked in specific regions of the non-Hermitian potential, in particular those with a vanishing value.
This phenomenology stems from the non-perturbative nature of the non-Hermitian potential, in stark contrast with the 
Hermitian case.

\section{Conclusion}
\label{sec:con}
We have shown how tunable local losses allow engineering topological modes in a
photonic system. In particular, we showed that both periodically engineered and
quasiperiodic loss profiles allow the creation of topological excitations at the edge
of the photonic array. In the quasiperiodic limit, we showed that a critical
localization takes place both in the presence of modulated losses, as well as
in generalized photonic arrays where both the local resonance frequency and the
loss are modulated. We showed that in the presence of second and
third-nearest-neighbor hopping, the localization transition takes place at
different modulation strengths for each frequency, leading to a photonic
spectrum featuring a mobility edge. Focusing on a topological regime featuring
zero modes, we addressed the robustness of the topological edge modes
in the presence of disorder, showing that resonance frequency disorder leads to
an energy splitting in the real energy, whereas loss disorder keeps states at
zero energy. Finally, we address the continuum limit
of the non-Hermitian model showing a similar emergence
of spectral mini-bands. 
Our results demonstrated that photonic arrays with periodically
modulated losses provide a flexible platform to engineer both topological modes
and criticality, making reconfigurable photonics a promising platform to
explore exotic non-Hermitian states of light.

\textbf{Acknowledgements}
We acknowledge the financial support of the 
Nokia Industrial Doctoral School in Quantum Technology.
JLL acknowledges the 
financial support from the
Academy of Finland Projects No. 331342 and No. 358088
and the Jane and Aatos Erkko Foundation.
AB-R acknowledges support by the National Science Foundation (NSF) (award ID 2328993).
ELP and JLL acknowledge the computational resources provided by
the Aalto Science-IT project.
\bibliography{biblio}

\begin{thebibliography}{63}%
\makeatletter
\providecommand \@ifxundefined [1]{%
 \@ifx{#1\undefined}
}%
\providecommand \@ifnum [1]{%
 \ifnum #1\expandafter \@firstoftwo
 \else \expandafter \@secondoftwo
 \fi
}%
\providecommand \@ifx [1]{%
 \ifx #1\expandafter \@firstoftwo
 \else \expandafter \@secondoftwo
 \fi
}%
\providecommand \natexlab [1]{#1}%
\providecommand \enquote  [1]{``#1''}%
\providecommand \bibnamefont  [1]{#1}%
\providecommand \bibfnamefont [1]{#1}%
\providecommand \citenamefont [1]{#1}%
\providecommand \href@noop [0]{\@secondoftwo}%
\providecommand \href [0]{\begingroup \@sanitize@url \@href}%
\providecommand \@href[1]{\@@startlink{#1}\@@href}%
\providecommand \@@href[1]{\endgroup#1\@@endlink}%
\providecommand \@sanitize@url [0]{\catcode `\\12\catcode `\$12\catcode
  `\&12\catcode `\#12\catcode `\^12\catcode `\_12\catcode `\%12\relax}%
\providecommand \@@startlink[1]{}%
\providecommand \@@endlink[0]{}%
\providecommand \url  [0]{\begingroup\@sanitize@url \@url }%
\providecommand \@url [1]{\endgroup\@href {#1}{\urlprefix }}%
\providecommand \urlprefix  [0]{URL }%
\providecommand \Eprint [0]{\href }%
\providecommand \doibase [0]{https://doi.org/}%
\providecommand \selectlanguage [0]{\@gobble}%
\providecommand \bibinfo  [0]{\@secondoftwo}%
\providecommand \bibfield  [0]{\@secondoftwo}%
\providecommand \translation [1]{[#1]}%
\providecommand \BibitemOpen [0]{}%
\providecommand \bibitemStop [0]{}%
\providecommand \bibitemNoStop [0]{.\EOS\space}%
\providecommand \EOS [0]{\spacefactor3000\relax}%
\providecommand \BibitemShut  [1]{\csname bibitem#1\endcsname}%
\let\auto@bib@innerbib\@empty
\bibitem [{\citenamefont {Haldane}(1988)}]{PhysRevLett.61.2015}%
  \BibitemOpen
  \bibfield  {author} {\bibinfo {author} {\bibfnamefont {F.~D.~M.}\
  \bibnamefont {Haldane}},\ }\bibfield  {title} {\bibinfo {title} {Model for a
  quantum hall effect without landau levels: Condensed-matter realization of
  the ``parity anomaly"},\ }\href {https://doi.org/10.1103/PhysRevLett.61.2015}
  {\bibfield  {journal} {\bibinfo  {journal} {Phys. Rev. Lett.}\ }\textbf
  {\bibinfo {volume} {61}},\ \bibinfo {pages} {2015} (\bibinfo {year}
  {1988})}\BibitemShut {NoStop}%
\bibitem [{\citenamefont {Qi}\ and\ \citenamefont
  {Zhang}(2011)}]{RevModPhys.83.1057}%
  \BibitemOpen
  \bibfield  {author} {\bibinfo {author} {\bibfnamefont {X.-L.}\ \bibnamefont
  {Qi}}\ and\ \bibinfo {author} {\bibfnamefont {S.-C.}\ \bibnamefont {Zhang}},\
  }\bibfield  {title} {\bibinfo {title} {Topological insulators and
  superconductors},\ }\href {https://doi.org/10.1103/RevModPhys.83.1057}
  {\bibfield  {journal} {\bibinfo  {journal} {Rev. Mod. Phys.}\ }\textbf
  {\bibinfo {volume} {83}},\ \bibinfo {pages} {1057} (\bibinfo {year}
  {2011})}\BibitemShut {NoStop}%
\bibitem [{\citenamefont {Hasan}\ and\ \citenamefont
  {Kane}(2010)}]{RevModPhys.82.3045}%
  \BibitemOpen
  \bibfield  {author} {\bibinfo {author} {\bibfnamefont {M.~Z.}\ \bibnamefont
  {Hasan}}\ and\ \bibinfo {author} {\bibfnamefont {C.~L.}\ \bibnamefont
  {Kane}},\ }\bibfield  {title} {\bibinfo {title} {Colloquium: Topological
  insulators},\ }\href {https://doi.org/10.1103/RevModPhys.82.3045} {\bibfield
  {journal} {\bibinfo  {journal} {Rev. Mod. Phys.}\ }\textbf {\bibinfo {volume}
  {82}},\ \bibinfo {pages} {3045} (\bibinfo {year} {2010})}\BibitemShut
  {NoStop}%
\bibitem [{\citenamefont {Huber}(2016)}]{Huber2016}%
  \BibitemOpen
  \bibfield  {author} {\bibinfo {author} {\bibfnamefont {S.~D.}\ \bibnamefont
  {Huber}},\ }\bibfield  {title} {\bibinfo {title} {Topological mechanics},\
  }\href {https://doi.org/10.1038/nphys3801} {\bibfield  {journal} {\bibinfo
  {journal} {Nature Physics}\ }\textbf {\bibinfo {volume} {12}},\ \bibinfo
  {pages} {621} (\bibinfo {year} {2016})}\BibitemShut {NoStop}%
\bibitem [{\citenamefont {Wang}\ \emph {et~al.}(2009)\citenamefont {Wang},
  \citenamefont {Chong}, \citenamefont {Joannopoulos},\ and\ \citenamefont
  {Soljačić}}]{Wang2009}%
  \BibitemOpen
  \bibfield  {author} {\bibinfo {author} {\bibfnamefont {Z.}~\bibnamefont
  {Wang}}, \bibinfo {author} {\bibfnamefont {Y.}~\bibnamefont {Chong}},
  \bibinfo {author} {\bibfnamefont {J.~D.}\ \bibnamefont {Joannopoulos}},\ and\
  \bibinfo {author} {\bibfnamefont {M.}~\bibnamefont {Soljačić}},\ }\bibfield
   {title} {\bibinfo {title} {Observation of unidirectional
  backscattering-immune topological electromagnetic states},\ }\href
  {https://doi.org/10.1038/nature08293} {\bibfield  {journal} {\bibinfo
  {journal} {Nature}\ }\textbf {\bibinfo {volume} {461}},\ \bibinfo {pages}
  {772–775} (\bibinfo {year} {2009})}\BibitemShut {NoStop}%
\bibitem [{\citenamefont {Khanikaev}\ \emph {et~al.}(2012)\citenamefont
  {Khanikaev}, \citenamefont {Mousavi}, \citenamefont {Tse}, \citenamefont
  {Kargarian}, \citenamefont {MacDonald},\ and\ \citenamefont
  {Shvets}}]{Khanikaev2012}%
  \BibitemOpen
  \bibfield  {author} {\bibinfo {author} {\bibfnamefont {A.~B.}\ \bibnamefont
  {Khanikaev}}, \bibinfo {author} {\bibfnamefont {S.~H.}\ \bibnamefont
  {Mousavi}}, \bibinfo {author} {\bibfnamefont {W.-K.}\ \bibnamefont {Tse}},
  \bibinfo {author} {\bibfnamefont {M.}~\bibnamefont {Kargarian}}, \bibinfo
  {author} {\bibfnamefont {A.~H.}\ \bibnamefont {MacDonald}},\ and\ \bibinfo
  {author} {\bibfnamefont {G.}~\bibnamefont {Shvets}},\ }\bibfield  {title}
  {\bibinfo {title} {Photonic topological insulators},\ }\href
  {https://doi.org/10.1038/nmat3520} {\bibfield  {journal} {\bibinfo  {journal}
  {Nature Materials}\ }\textbf {\bibinfo {volume} {12}},\ \bibinfo {pages}
  {233} (\bibinfo {year} {2012})}\BibitemShut {NoStop}%
\bibitem [{\citenamefont {Jotzu}\ \emph {et~al.}(2014)\citenamefont {Jotzu},
  \citenamefont {Messer}, \citenamefont {Desbuquois}, \citenamefont {Lebrat},
  \citenamefont {Uehlinger}, \citenamefont {Greif},\ and\ \citenamefont
  {Esslinger}}]{Jotzu2014}%
  \BibitemOpen
  \bibfield  {author} {\bibinfo {author} {\bibfnamefont {G.}~\bibnamefont
  {Jotzu}}, \bibinfo {author} {\bibfnamefont {M.}~\bibnamefont {Messer}},
  \bibinfo {author} {\bibfnamefont {R.}~\bibnamefont {Desbuquois}}, \bibinfo
  {author} {\bibfnamefont {M.}~\bibnamefont {Lebrat}}, \bibinfo {author}
  {\bibfnamefont {T.}~\bibnamefont {Uehlinger}}, \bibinfo {author}
  {\bibfnamefont {D.}~\bibnamefont {Greif}},\ and\ \bibinfo {author}
  {\bibfnamefont {T.}~\bibnamefont {Esslinger}},\ }\bibfield  {title} {\bibinfo
  {title} {Experimental realization of the topological haldane model with
  ultracold fermions},\ }\href {https://doi.org/10.1038/nature13915} {\bibfield
   {journal} {\bibinfo  {journal} {Nature}\ }\textbf {\bibinfo {volume}
  {515}},\ \bibinfo {pages} {237} (\bibinfo {year} {2014})}\BibitemShut
  {NoStop}%
\bibitem [{\citenamefont {Ozawa}\ \emph {et~al.}(2019)\citenamefont {Ozawa},
  \citenamefont {Price}, \citenamefont {Amo}, \citenamefont {Goldman},
  \citenamefont {Hafezi}, \citenamefont {Lu}, \citenamefont {Rechtsman},
  \citenamefont {Schuster}, \citenamefont {Simon}, \citenamefont {Zilberberg},\
  and\ \citenamefont {Carusotto}}]{RevModPhys.91.015006}%
  \BibitemOpen
  \bibfield  {author} {\bibinfo {author} {\bibfnamefont {T.}~\bibnamefont
  {Ozawa}}, \bibinfo {author} {\bibfnamefont {H.~M.}\ \bibnamefont {Price}},
  \bibinfo {author} {\bibfnamefont {A.}~\bibnamefont {Amo}}, \bibinfo {author}
  {\bibfnamefont {N.}~\bibnamefont {Goldman}}, \bibinfo {author} {\bibfnamefont
  {M.}~\bibnamefont {Hafezi}}, \bibinfo {author} {\bibfnamefont
  {L.}~\bibnamefont {Lu}}, \bibinfo {author} {\bibfnamefont {M.~C.}\
  \bibnamefont {Rechtsman}}, \bibinfo {author} {\bibfnamefont {D.}~\bibnamefont
  {Schuster}}, \bibinfo {author} {\bibfnamefont {J.}~\bibnamefont {Simon}},
  \bibinfo {author} {\bibfnamefont {O.}~\bibnamefont {Zilberberg}},\ and\
  \bibinfo {author} {\bibfnamefont {I.}~\bibnamefont {Carusotto}},\ }\bibfield
  {title} {\bibinfo {title} {Topological photonics},\ }\href
  {https://doi.org/10.1103/RevModPhys.91.015006} {\bibfield  {journal}
  {\bibinfo  {journal} {Rev. Mod. Phys.}\ }\textbf {\bibinfo {volume} {91}},\
  \bibinfo {pages} {015006} (\bibinfo {year} {2019})}\BibitemShut {NoStop}%
\bibitem [{\citenamefont {Price}\ \emph {et~al.}(2022)\citenamefont {Price},
  \citenamefont {Chong}, \citenamefont {Khanikaev}, \citenamefont {Schomerus},
  \citenamefont {Maczewsky}, \citenamefont {Kremer}, \citenamefont {Heinrich},
  \citenamefont {Szameit}, \citenamefont {Zilberberg}, \citenamefont {Yang},
  \citenamefont {Zhang}, \citenamefont {Alù}, \citenamefont {Thomale},
  \citenamefont {Carusotto}, \citenamefont {St-Jean}, \citenamefont {Amo},
  \citenamefont {Dutt}, \citenamefont {Yuan}, \citenamefont {Fan},
  \citenamefont {Yin}, \citenamefont {Peng}, \citenamefont {Ozawa},\ and\
  \citenamefont {Blanco-Redondo}}]{price2022}%
  \BibitemOpen
  \bibfield  {author} {\bibinfo {author} {\bibfnamefont {H.}~\bibnamefont
  {Price}}, \bibinfo {author} {\bibfnamefont {Y.}~\bibnamefont {Chong}},
  \bibinfo {author} {\bibfnamefont {A.}~\bibnamefont {Khanikaev}}, \bibinfo
  {author} {\bibfnamefont {H.}~\bibnamefont {Schomerus}}, \bibinfo {author}
  {\bibfnamefont {L.~J.}\ \bibnamefont {Maczewsky}}, \bibinfo {author}
  {\bibfnamefont {M.}~\bibnamefont {Kremer}}, \bibinfo {author} {\bibfnamefont
  {M.}~\bibnamefont {Heinrich}}, \bibinfo {author} {\bibfnamefont
  {A.}~\bibnamefont {Szameit}}, \bibinfo {author} {\bibfnamefont
  {O.}~\bibnamefont {Zilberberg}}, \bibinfo {author} {\bibfnamefont
  {Y.}~\bibnamefont {Yang}}, \bibinfo {author} {\bibfnamefont {B.}~\bibnamefont
  {Zhang}}, \bibinfo {author} {\bibfnamefont {A.}~\bibnamefont {Alù}},
  \bibinfo {author} {\bibfnamefont {R.}~\bibnamefont {Thomale}}, \bibinfo
  {author} {\bibfnamefont {I.}~\bibnamefont {Carusotto}}, \bibinfo {author}
  {\bibfnamefont {P.}~\bibnamefont {St-Jean}}, \bibinfo {author} {\bibfnamefont
  {A.}~\bibnamefont {Amo}}, \bibinfo {author} {\bibfnamefont {A.}~\bibnamefont
  {Dutt}}, \bibinfo {author} {\bibfnamefont {L.}~\bibnamefont {Yuan}}, \bibinfo
  {author} {\bibfnamefont {S.}~\bibnamefont {Fan}}, \bibinfo {author}
  {\bibfnamefont {X.}~\bibnamefont {Yin}}, \bibinfo {author} {\bibfnamefont
  {C.}~\bibnamefont {Peng}}, \bibinfo {author} {\bibfnamefont {T.}~\bibnamefont
  {Ozawa}},\ and\ \bibinfo {author} {\bibfnamefont {A.}~\bibnamefont
  {Blanco-Redondo}},\ }\bibfield  {title} {\bibinfo {title} {Roadmap on
  topological photonics},\ }\href {https://doi.org/10.1088/2515-7647/ac4ee4}
  {\bibfield  {journal} {\bibinfo  {journal} {Journal of Physics: Photonics}\
  }\textbf {\bibinfo {volume} {4}},\ \bibinfo {pages} {032501} (\bibinfo {year}
  {2022})}\BibitemShut {NoStop}%
\bibitem [{\citenamefont {Bandres}\ \emph {et~al.}(2018)\citenamefont
  {Bandres}, \citenamefont {Wittek}, \citenamefont {Harari}, \citenamefont
  {Parto}, \citenamefont {Ren}, \citenamefont {Segev}, \citenamefont
  {Christodoulides},\ and\ \citenamefont {Khajavikhan}}]{Bandres2018}%
  \BibitemOpen
  \bibfield  {author} {\bibinfo {author} {\bibfnamefont {M.~A.}\ \bibnamefont
  {Bandres}}, \bibinfo {author} {\bibfnamefont {S.}~\bibnamefont {Wittek}},
  \bibinfo {author} {\bibfnamefont {G.}~\bibnamefont {Harari}}, \bibinfo
  {author} {\bibfnamefont {M.}~\bibnamefont {Parto}}, \bibinfo {author}
  {\bibfnamefont {J.}~\bibnamefont {Ren}}, \bibinfo {author} {\bibfnamefont
  {M.}~\bibnamefont {Segev}}, \bibinfo {author} {\bibfnamefont {D.~N.}\
  \bibnamefont {Christodoulides}},\ and\ \bibinfo {author} {\bibfnamefont
  {M.}~\bibnamefont {Khajavikhan}},\ }\bibfield  {title} {\bibinfo {title}
  {Topological insulator laser: Experiments},\ }\bibfield  {journal} {\bibinfo
  {journal} {Science}\ }\textbf {\bibinfo {volume} {359}},\ \href
  {https://doi.org/10.1126/science.aar4005} {10.1126/science.aar4005} (\bibinfo
  {year} {2018})\BibitemShut {NoStop}%
\bibitem [{\citenamefont {Contractor}\ \emph {et~al.}(2022)\citenamefont
  {Contractor}, \citenamefont {Noh}, \citenamefont {Redjem}, \citenamefont
  {Qarony}, \citenamefont {Martin}, \citenamefont {Dhuey}, \citenamefont
  {Schwartzberg},\ and\ \citenamefont {Kant{\'{e}}}}]{Contractor2022}%
  \BibitemOpen
  \bibfield  {author} {\bibinfo {author} {\bibfnamefont {R.}~\bibnamefont
  {Contractor}}, \bibinfo {author} {\bibfnamefont {W.}~\bibnamefont {Noh}},
  \bibinfo {author} {\bibfnamefont {W.}~\bibnamefont {Redjem}}, \bibinfo
  {author} {\bibfnamefont {W.}~\bibnamefont {Qarony}}, \bibinfo {author}
  {\bibfnamefont {E.}~\bibnamefont {Martin}}, \bibinfo {author} {\bibfnamefont
  {S.}~\bibnamefont {Dhuey}}, \bibinfo {author} {\bibfnamefont
  {A.}~\bibnamefont {Schwartzberg}},\ and\ \bibinfo {author} {\bibfnamefont
  {B.}~\bibnamefont {Kant{\'{e}}}},\ }\bibfield  {title} {\bibinfo {title}
  {Scalable single-mode surface-emitting laser via open-dirac singularities},\
  }\href {https://doi.org/10.1038/s41586-022-05021-4} {\bibfield  {journal}
  {\bibinfo  {journal} {Nature}\ }\textbf {\bibinfo {volume} {608}},\ \bibinfo
  {pages} {692} (\bibinfo {year} {2022})}\BibitemShut {NoStop}%
\bibitem [{\citenamefont {Blanco-Redondo}\ \emph {et~al.}(2018)\citenamefont
  {Blanco-Redondo}, \citenamefont {Bell}, \citenamefont {Oren}, \citenamefont
  {Eggleton},\ and\ \citenamefont {Segev}}]{BlancoRedondo2018}%
  \BibitemOpen
  \bibfield  {author} {\bibinfo {author} {\bibfnamefont {A.}~\bibnamefont
  {Blanco-Redondo}}, \bibinfo {author} {\bibfnamefont {B.}~\bibnamefont
  {Bell}}, \bibinfo {author} {\bibfnamefont {D.}~\bibnamefont {Oren}}, \bibinfo
  {author} {\bibfnamefont {B.~J.}\ \bibnamefont {Eggleton}},\ and\ \bibinfo
  {author} {\bibfnamefont {M.}~\bibnamefont {Segev}},\ }\bibfield  {title}
  {\bibinfo {title} {Topological protection of biphoton states},\ }\href
  {https://doi.org/10.1126/science.aau4296} {\bibfield  {journal} {\bibinfo
  {journal} {Science}\ }\textbf {\bibinfo {volume} {362}},\ \bibinfo {pages}
  {568} (\bibinfo {year} {2018})}\BibitemShut {NoStop}%
\bibitem [{\citenamefont {Mittal}\ \emph {et~al.}(2018)\citenamefont {Mittal},
  \citenamefont {Goldschmidt},\ and\ \citenamefont {Hafezi}}]{Mittal2018}%
  \BibitemOpen
  \bibfield  {author} {\bibinfo {author} {\bibfnamefont {S.}~\bibnamefont
  {Mittal}}, \bibinfo {author} {\bibfnamefont {E.~A.}\ \bibnamefont
  {Goldschmidt}},\ and\ \bibinfo {author} {\bibfnamefont {M.}~\bibnamefont
  {Hafezi}},\ }\bibfield  {title} {\bibinfo {title} {A topological source of
  quantum light},\ }\href {https://doi.org/10.1038/s41586-018-0478-3}
  {\bibfield  {journal} {\bibinfo  {journal} {Nature}\ }\textbf {\bibinfo
  {volume} {561}},\ \bibinfo {pages} {502} (\bibinfo {year}
  {2018})}\BibitemShut {NoStop}%
\bibitem [{\citenamefont {Wang}\ \emph {et~al.}(2019)\citenamefont {Wang},
  \citenamefont {Doyle}, \citenamefont {Bell}, \citenamefont {Collins},
  \citenamefont {Magi}, \citenamefont {Eggleton}, \citenamefont {Segev},\ and\
  \citenamefont {Blanco-Redondo}}]{Wang2019}%
  \BibitemOpen
  \bibfield  {author} {\bibinfo {author} {\bibfnamefont {M.}~\bibnamefont
  {Wang}}, \bibinfo {author} {\bibfnamefont {C.}~\bibnamefont {Doyle}},
  \bibinfo {author} {\bibfnamefont {B.}~\bibnamefont {Bell}}, \bibinfo {author}
  {\bibfnamefont {M.~J.}\ \bibnamefont {Collins}}, \bibinfo {author}
  {\bibfnamefont {E.}~\bibnamefont {Magi}}, \bibinfo {author} {\bibfnamefont
  {B.~J.}\ \bibnamefont {Eggleton}}, \bibinfo {author} {\bibfnamefont
  {M.}~\bibnamefont {Segev}},\ and\ \bibinfo {author} {\bibfnamefont
  {A.}~\bibnamefont {Blanco-Redondo}},\ }\bibfield  {title} {\bibinfo {title}
  {Topologically protected entangled photonic states},\ }\href
  {https://doi.org/10.1515/nanoph-2019-0058} {\bibfield  {journal} {\bibinfo
  {journal} {Nanophotonics}\ }\textbf {\bibinfo {volume} {8}},\ \bibinfo
  {pages} {1327} (\bibinfo {year} {2019})}\BibitemShut {NoStop}%
\bibitem [{\citenamefont {Sahlberg}\ \emph {et~al.}(2020)\citenamefont
  {Sahlberg}, \citenamefont {Weststr\"om}, \citenamefont {P\"oyh\"onen},\ and\
  \citenamefont {Ojanen}}]{PhysRevResearch.2.013053}%
  \BibitemOpen
  \bibfield  {author} {\bibinfo {author} {\bibfnamefont {I.}~\bibnamefont
  {Sahlberg}}, \bibinfo {author} {\bibfnamefont {A.}~\bibnamefont
  {Weststr\"om}}, \bibinfo {author} {\bibfnamefont {K.}~\bibnamefont
  {P\"oyh\"onen}},\ and\ \bibinfo {author} {\bibfnamefont {T.}~\bibnamefont
  {Ojanen}},\ }\bibfield  {title} {\bibinfo {title} {Topological phase
  transitions in glassy quantum matter},\ }\href
  {https://doi.org/10.1103/PhysRevResearch.2.013053} {\bibfield  {journal}
  {\bibinfo  {journal} {Phys. Rev. Res.}\ }\textbf {\bibinfo {volume} {2}},\
  \bibinfo {pages} {013053} (\bibinfo {year} {2020})}\BibitemShut {NoStop}%
\bibitem [{\citenamefont {P\"{o}yh\"{o}nen}\ \emph {et~al.}(2018)\citenamefont
  {P\"{o}yh\"{o}nen}, \citenamefont {Sahlberg}, \citenamefont {Weststr\"{o}m},\
  and\ \citenamefont {Ojanen}}]{Pyhnen2018}%
  \BibitemOpen
  \bibfield  {author} {\bibinfo {author} {\bibfnamefont {K.}~\bibnamefont
  {P\"{o}yh\"{o}nen}}, \bibinfo {author} {\bibfnamefont {I.}~\bibnamefont
  {Sahlberg}}, \bibinfo {author} {\bibfnamefont {A.}~\bibnamefont
  {Weststr\"{o}m}},\ and\ \bibinfo {author} {\bibfnamefont {T.}~\bibnamefont
  {Ojanen}},\ }\bibfield  {title} {\bibinfo {title} {Amorphous topological
  superconductivity in a shiba glass},\ }\bibfield  {journal} {\bibinfo
  {journal} {Nature Communications}\ }\textbf {\bibinfo {volume} {9}},\ \href
  {https://doi.org/10.1038/s41467-018-04532-x} {10.1038/s41467-018-04532-x}
  (\bibinfo {year} {2018})\BibitemShut {NoStop}%
\bibitem [{\citenamefont {Marsal}\ \emph {et~al.}(2020)\citenamefont {Marsal},
  \citenamefont {Varjas},\ and\ \citenamefont {Grushin}}]{Marsal2020}%
  \BibitemOpen
  \bibfield  {author} {\bibinfo {author} {\bibfnamefont {Q.}~\bibnamefont
  {Marsal}}, \bibinfo {author} {\bibfnamefont {D.}~\bibnamefont {Varjas}},\
  and\ \bibinfo {author} {\bibfnamefont {A.~G.}\ \bibnamefont {Grushin}},\
  }\bibfield  {title} {\bibinfo {title} {Topological weaire{\textendash}thorpe
  models of amorphous matter},\ }\href
  {https://doi.org/10.1073/pnas.2007384117} {\bibfield  {journal} {\bibinfo
  {journal} {Proceedings of the National Academy of Sciences}\ }\textbf
  {\bibinfo {volume} {117}},\ \bibinfo {pages} {30260} (\bibinfo {year}
  {2020})}\BibitemShut {NoStop}%
\bibitem [{\citenamefont {Shechtman}\ \emph {et~al.}(1984)\citenamefont
  {Shechtman}, \citenamefont {Blech}, \citenamefont {Gratias},\ and\
  \citenamefont {Cahn}}]{PhysRevLett.53.1951}%
  \BibitemOpen
  \bibfield  {author} {\bibinfo {author} {\bibfnamefont {D.}~\bibnamefont
  {Shechtman}}, \bibinfo {author} {\bibfnamefont {I.}~\bibnamefont {Blech}},
  \bibinfo {author} {\bibfnamefont {D.}~\bibnamefont {Gratias}},\ and\ \bibinfo
  {author} {\bibfnamefont {J.~W.}\ \bibnamefont {Cahn}},\ }\bibfield  {title}
  {\bibinfo {title} {Metallic phase with long-range orientational order and no
  translational symmetry},\ }\href
  {https://doi.org/10.1103/PhysRevLett.53.1951} {\bibfield  {journal} {\bibinfo
   {journal} {Phys. Rev. Lett.}\ }\textbf {\bibinfo {volume} {53}},\ \bibinfo
  {pages} {1951} (\bibinfo {year} {1984})}\BibitemShut {NoStop}%
\bibitem [{\citenamefont {Levine}\ and\ \citenamefont
  {Steinhardt}(1984)}]{PhysRevLett.53.2477}%
  \BibitemOpen
  \bibfield  {author} {\bibinfo {author} {\bibfnamefont {D.}~\bibnamefont
  {Levine}}\ and\ \bibinfo {author} {\bibfnamefont {P.~J.}\ \bibnamefont
  {Steinhardt}},\ }\bibfield  {title} {\bibinfo {title} {Quasicrystals: A new
  class of ordered structures},\ }\href
  {https://doi.org/10.1103/PhysRevLett.53.2477} {\bibfield  {journal} {\bibinfo
   {journal} {Phys. Rev. Lett.}\ }\textbf {\bibinfo {volume} {53}},\ \bibinfo
  {pages} {2477} (\bibinfo {year} {1984})}\BibitemShut {NoStop}%
\bibitem [{\citenamefont {Aubry}\ and\ \citenamefont {André}(1980)}]{aubry}%
  \BibitemOpen
  \bibfield  {author} {\bibinfo {author} {\bibfnamefont {S.}~\bibnamefont
  {Aubry}}\ and\ \bibinfo {author} {\bibfnamefont {G.}~\bibnamefont {André}},\
  }\bibfield  {title} {\bibinfo {title} {Analyticity breaking and anderson
  localization in incommensurate lattices},\ }\href@noop {} {\bibfield
  {journal} {\bibinfo  {journal} {Proceedings, VIII International Colloquium on
  Group-Theoretical Methods in Physics}\ }\textbf {\bibinfo {volume} {3}}
  (\bibinfo {year} {1980})}\BibitemShut {NoStop}%
\bibitem [{\citenamefont {Price}\ \emph {et~al.}(2015)\citenamefont {Price},
  \citenamefont {Zilberberg}, \citenamefont {Ozawa}, \citenamefont
  {Carusotto},\ and\ \citenamefont {Goldman}}]{PhysRevLett.115.195303}%
  \BibitemOpen
  \bibfield  {author} {\bibinfo {author} {\bibfnamefont {H.~M.}\ \bibnamefont
  {Price}}, \bibinfo {author} {\bibfnamefont {O.}~\bibnamefont {Zilberberg}},
  \bibinfo {author} {\bibfnamefont {T.}~\bibnamefont {Ozawa}}, \bibinfo
  {author} {\bibfnamefont {I.}~\bibnamefont {Carusotto}},\ and\ \bibinfo
  {author} {\bibfnamefont {N.}~\bibnamefont {Goldman}},\ }\bibfield  {title}
  {\bibinfo {title} {Four-dimensional quantum hall effect with ultracold
  atoms},\ }\href {https://doi.org/10.1103/PhysRevLett.115.195303} {\bibfield
  {journal} {\bibinfo  {journal} {Phys. Rev. Lett.}\ }\textbf {\bibinfo
  {volume} {115}},\ \bibinfo {pages} {195303} (\bibinfo {year}
  {2015})}\BibitemShut {NoStop}%
\bibitem [{\citenamefont {Kraus}\ \emph {et~al.}(2012)\citenamefont {Kraus},
  \citenamefont {Lahini}, \citenamefont {Ringel}, \citenamefont {Verbin},\ and\
  \citenamefont {Zilberberg}}]{PhysRevLett.109.106402}%
  \BibitemOpen
  \bibfield  {author} {\bibinfo {author} {\bibfnamefont {Y.~E.}\ \bibnamefont
  {Kraus}}, \bibinfo {author} {\bibfnamefont {Y.}~\bibnamefont {Lahini}},
  \bibinfo {author} {\bibfnamefont {Z.}~\bibnamefont {Ringel}}, \bibinfo
  {author} {\bibfnamefont {M.}~\bibnamefont {Verbin}},\ and\ \bibinfo {author}
  {\bibfnamefont {O.}~\bibnamefont {Zilberberg}},\ }\bibfield  {title}
  {\bibinfo {title} {Topological states and adiabatic pumping in
  quasicrystals},\ }\href {https://doi.org/10.1103/PhysRevLett.109.106402}
  {\bibfield  {journal} {\bibinfo  {journal} {Phys. Rev. Lett.}\ }\textbf
  {\bibinfo {volume} {109}},\ \bibinfo {pages} {106402} (\bibinfo {year}
  {2012})}\BibitemShut {NoStop}%
\bibitem [{\citenamefont {Petrides}\ and\ \citenamefont
  {Zilberberg}(2020)}]{PhysRevResearch.2.022049}%
  \BibitemOpen
  \bibfield  {author} {\bibinfo {author} {\bibfnamefont {I.}~\bibnamefont
  {Petrides}}\ and\ \bibinfo {author} {\bibfnamefont {O.}~\bibnamefont
  {Zilberberg}},\ }\bibfield  {title} {\bibinfo {title} {Higher-order
  topological insulators, topological pumps and the quantum hall effect in high
  dimensions},\ }\href {https://doi.org/10.1103/PhysRevResearch.2.022049}
  {\bibfield  {journal} {\bibinfo  {journal} {Phys. Rev. Res.}\ }\textbf
  {\bibinfo {volume} {2}},\ \bibinfo {pages} {022049} (\bibinfo {year}
  {2020})}\BibitemShut {NoStop}%
\bibitem [{\citenamefont {Petrides}\ \emph {et~al.}(2018)\citenamefont
  {Petrides}, \citenamefont {Price},\ and\ \citenamefont
  {Zilberberg}}]{PhysRevB.98.125431}%
  \BibitemOpen
  \bibfield  {author} {\bibinfo {author} {\bibfnamefont {I.}~\bibnamefont
  {Petrides}}, \bibinfo {author} {\bibfnamefont {H.~M.}\ \bibnamefont
  {Price}},\ and\ \bibinfo {author} {\bibfnamefont {O.}~\bibnamefont
  {Zilberberg}},\ }\bibfield  {title} {\bibinfo {title} {Six-dimensional
  quantum hall effect and three-dimensional topological pumps},\ }\href
  {https://doi.org/10.1103/PhysRevB.98.125431} {\bibfield  {journal} {\bibinfo
  {journal} {Phys. Rev. B}\ }\textbf {\bibinfo {volume} {98}},\ \bibinfo
  {pages} {125431} (\bibinfo {year} {2018})}\BibitemShut {NoStop}%
\bibitem [{\citenamefont {Verbin}\ \emph {et~al.}(2015)\citenamefont {Verbin},
  \citenamefont {Zilberberg}, \citenamefont {Lahini}, \citenamefont {Kraus},\
  and\ \citenamefont {Silberberg}}]{PhysRevB.91.064201}%
  \BibitemOpen
  \bibfield  {author} {\bibinfo {author} {\bibfnamefont {M.}~\bibnamefont
  {Verbin}}, \bibinfo {author} {\bibfnamefont {O.}~\bibnamefont {Zilberberg}},
  \bibinfo {author} {\bibfnamefont {Y.}~\bibnamefont {Lahini}}, \bibinfo
  {author} {\bibfnamefont {Y.~E.}\ \bibnamefont {Kraus}},\ and\ \bibinfo
  {author} {\bibfnamefont {Y.}~\bibnamefont {Silberberg}},\ }\bibfield  {title}
  {\bibinfo {title} {Topological pumping over a photonic fibonacci
  quasicrystal},\ }\href {https://doi.org/10.1103/PhysRevB.91.064201}
  {\bibfield  {journal} {\bibinfo  {journal} {Phys. Rev. B}\ }\textbf {\bibinfo
  {volume} {91}},\ \bibinfo {pages} {064201} (\bibinfo {year}
  {2015})}\BibitemShut {NoStop}%
\bibitem [{\citenamefont {Zilberberg}(2021)}]{Zilberberg2021}%
  \BibitemOpen
  \bibfield  {author} {\bibinfo {author} {\bibfnamefont {O.}~\bibnamefont
  {Zilberberg}},\ }\bibfield  {title} {\bibinfo {title} {Topology in
  quasicrystals [invited]},\ }\href {https://doi.org/10.1364/ome.416552}
  {\bibfield  {journal} {\bibinfo  {journal} {Optical Materials Express}\
  }\textbf {\bibinfo {volume} {11}},\ \bibinfo {pages} {1143} (\bibinfo {year}
  {2021})}\BibitemShut {NoStop}%
\bibitem [{\citenamefont {Goblot}\ \emph {et~al.}(2020)\citenamefont {Goblot},
  \citenamefont {{\v{S}}trkalj}, \citenamefont {Pernet}, \citenamefont {Lado},
  \citenamefont {Dorow}, \citenamefont {Lema{\^{\i}}tre}, \citenamefont
  {Gratiet}, \citenamefont {Harouri}, \citenamefont {Sagnes}, \citenamefont
  {Ravets}, \citenamefont {Amo}, \citenamefont {Bloch},\ and\ \citenamefont
  {Zilberberg}}]{Goblot2020}%
  \BibitemOpen
  \bibfield  {author} {\bibinfo {author} {\bibfnamefont {V.}~\bibnamefont
  {Goblot}}, \bibinfo {author} {\bibfnamefont {A.}~\bibnamefont
  {{\v{S}}trkalj}}, \bibinfo {author} {\bibfnamefont {N.}~\bibnamefont
  {Pernet}}, \bibinfo {author} {\bibfnamefont {J.~L.}\ \bibnamefont {Lado}},
  \bibinfo {author} {\bibfnamefont {C.}~\bibnamefont {Dorow}}, \bibinfo
  {author} {\bibfnamefont {A.}~\bibnamefont {Lema{\^{\i}}tre}}, \bibinfo
  {author} {\bibfnamefont {L.~L.}\ \bibnamefont {Gratiet}}, \bibinfo {author}
  {\bibfnamefont {A.}~\bibnamefont {Harouri}}, \bibinfo {author} {\bibfnamefont
  {I.}~\bibnamefont {Sagnes}}, \bibinfo {author} {\bibfnamefont
  {S.}~\bibnamefont {Ravets}}, \bibinfo {author} {\bibfnamefont
  {A.}~\bibnamefont {Amo}}, \bibinfo {author} {\bibfnamefont {J.}~\bibnamefont
  {Bloch}},\ and\ \bibinfo {author} {\bibfnamefont {O.}~\bibnamefont
  {Zilberberg}},\ }\bibfield  {title} {\bibinfo {title} {Emergence of
  criticality through a cascade of delocalization transitions in quasiperiodic
  chains},\ }\href {https://doi.org/10.1038/s41567-020-0908-7} {\bibfield
  {journal} {\bibinfo  {journal} {Nature Physics}\ }\textbf {\bibinfo {volume}
  {16}},\ \bibinfo {pages} {832} (\bibinfo {year} {2020})}\BibitemShut
  {NoStop}%
\bibitem [{\citenamefont {Zilberberg}\ \emph {et~al.}(2018)\citenamefont
  {Zilberberg}, \citenamefont {Huang}, \citenamefont {Guglielmon},
  \citenamefont {Wang}, \citenamefont {Chen}, \citenamefont {Kraus},\ and\
  \citenamefont {Rechtsman}}]{Zilberberg2018}%
  \BibitemOpen
  \bibfield  {author} {\bibinfo {author} {\bibfnamefont {O.}~\bibnamefont
  {Zilberberg}}, \bibinfo {author} {\bibfnamefont {S.}~\bibnamefont {Huang}},
  \bibinfo {author} {\bibfnamefont {J.}~\bibnamefont {Guglielmon}}, \bibinfo
  {author} {\bibfnamefont {M.}~\bibnamefont {Wang}}, \bibinfo {author}
  {\bibfnamefont {K.~P.}\ \bibnamefont {Chen}}, \bibinfo {author}
  {\bibfnamefont {Y.~E.}\ \bibnamefont {Kraus}},\ and\ \bibinfo {author}
  {\bibfnamefont {M.~C.}\ \bibnamefont {Rechtsman}},\ }\bibfield  {title}
  {\bibinfo {title} {Photonic topological boundary pumping as a probe of 4d
  quantum hall physics},\ }\href {https://doi.org/10.1038/nature25011}
  {\bibfield  {journal} {\bibinfo  {journal} {Nature}\ }\textbf {\bibinfo
  {volume} {553}},\ \bibinfo {pages} {59} (\bibinfo {year} {2018})}\BibitemShut
  {NoStop}%
\bibitem [{\citenamefont {St-Jean}\ \emph {et~al.}(2017)\citenamefont
  {St-Jean}, \citenamefont {Goblot}, \citenamefont {Galopin}, \citenamefont
  {Lema{\^{\i}}tre}, \citenamefont {Ozawa}, \citenamefont {Gratiet},
  \citenamefont {Sagnes}, \citenamefont {Bloch},\ and\ \citenamefont
  {Amo}}]{StJean2017}%
  \BibitemOpen
  \bibfield  {author} {\bibinfo {author} {\bibfnamefont {P.}~\bibnamefont
  {St-Jean}}, \bibinfo {author} {\bibfnamefont {V.}~\bibnamefont {Goblot}},
  \bibinfo {author} {\bibfnamefont {E.}~\bibnamefont {Galopin}}, \bibinfo
  {author} {\bibfnamefont {A.}~\bibnamefont {Lema{\^{\i}}tre}}, \bibinfo
  {author} {\bibfnamefont {T.}~\bibnamefont {Ozawa}}, \bibinfo {author}
  {\bibfnamefont {L.~L.}\ \bibnamefont {Gratiet}}, \bibinfo {author}
  {\bibfnamefont {I.}~\bibnamefont {Sagnes}}, \bibinfo {author} {\bibfnamefont
  {J.}~\bibnamefont {Bloch}},\ and\ \bibinfo {author} {\bibfnamefont
  {A.}~\bibnamefont {Amo}},\ }\bibfield  {title} {\bibinfo {title} {Lasing in
  topological edge states of a one-dimensional lattice},\ }\href
  {https://doi.org/10.1038/s41566-017-0006-2} {\bibfield  {journal} {\bibinfo
  {journal} {Nature Photonics}\ }\textbf {\bibinfo {volume} {11}},\ \bibinfo
  {pages} {651} (\bibinfo {year} {2017})}\BibitemShut {NoStop}%
\bibitem [{\citenamefont {Okuma}\ and\ \citenamefont {Sato}(2023)}]{Okuma2023}%
  \BibitemOpen
  \bibfield  {author} {\bibinfo {author} {\bibfnamefont {N.}~\bibnamefont
  {Okuma}}\ and\ \bibinfo {author} {\bibfnamefont {M.}~\bibnamefont {Sato}},\
  }\bibfield  {title} {\bibinfo {title} {Non-hermitian topological phenomena: A
  review},\ }\href {https://doi.org/10.1146/annurev-conmatphys-040521-033133}
  {\bibfield  {journal} {\bibinfo  {journal} {Annual Review of Condensed Matter
  Physics}\ }\textbf {\bibinfo {volume} {14}},\ \bibinfo {pages} {83} (\bibinfo
  {year} {2023})}\BibitemShut {NoStop}%
\bibitem [{\citenamefont {Bergholtz}\ \emph {et~al.}(2021)\citenamefont
  {Bergholtz}, \citenamefont {Budich},\ and\ \citenamefont
  {Kunst}}]{RevModPhys.93.015005}%
  \BibitemOpen
  \bibfield  {author} {\bibinfo {author} {\bibfnamefont {E.~J.}\ \bibnamefont
  {Bergholtz}}, \bibinfo {author} {\bibfnamefont {J.~C.}\ \bibnamefont
  {Budich}},\ and\ \bibinfo {author} {\bibfnamefont {F.~K.}\ \bibnamefont
  {Kunst}},\ }\bibfield  {title} {\bibinfo {title} {Exceptional topology of
  non-hermitian systems},\ }\href
  {https://doi.org/10.1103/RevModPhys.93.015005} {\bibfield  {journal}
  {\bibinfo  {journal} {Rev. Mod. Phys.}\ }\textbf {\bibinfo {volume} {93}},\
  \bibinfo {pages} {015005} (\bibinfo {year} {2021})}\BibitemShut {NoStop}%
\bibitem [{\citenamefont {Kunst}\ \emph {et~al.}(2018)\citenamefont {Kunst},
  \citenamefont {Edvardsson}, \citenamefont {Budich},\ and\ \citenamefont
  {Bergholtz}}]{PhysRevLett.121.026808}%
  \BibitemOpen
  \bibfield  {author} {\bibinfo {author} {\bibfnamefont {F.~K.}\ \bibnamefont
  {Kunst}}, \bibinfo {author} {\bibfnamefont {E.}~\bibnamefont {Edvardsson}},
  \bibinfo {author} {\bibfnamefont {J.~C.}\ \bibnamefont {Budich}},\ and\
  \bibinfo {author} {\bibfnamefont {E.~J.}\ \bibnamefont {Bergholtz}},\
  }\bibfield  {title} {\bibinfo {title} {Biorthogonal bulk-boundary
  correspondence in non-hermitian systems},\ }\href
  {https://doi.org/10.1103/PhysRevLett.121.026808} {\bibfield  {journal}
  {\bibinfo  {journal} {Phys. Rev. Lett.}\ }\textbf {\bibinfo {volume} {121}},\
  \bibinfo {pages} {026808} (\bibinfo {year} {2018})}\BibitemShut {NoStop}%
\bibitem [{\citenamefont {Esaki}\ \emph {et~al.}(2011)\citenamefont {Esaki},
  \citenamefont {Sato}, \citenamefont {Hasebe},\ and\ \citenamefont
  {Kohmoto}}]{PhysRevB.84.205128}%
  \BibitemOpen
  \bibfield  {author} {\bibinfo {author} {\bibfnamefont {K.}~\bibnamefont
  {Esaki}}, \bibinfo {author} {\bibfnamefont {M.}~\bibnamefont {Sato}},
  \bibinfo {author} {\bibfnamefont {K.}~\bibnamefont {Hasebe}},\ and\ \bibinfo
  {author} {\bibfnamefont {M.}~\bibnamefont {Kohmoto}},\ }\bibfield  {title}
  {\bibinfo {title} {Edge states and topological phases in non-hermitian
  systems},\ }\href {https://doi.org/10.1103/PhysRevB.84.205128} {\bibfield
  {journal} {\bibinfo  {journal} {Phys. Rev. B}\ }\textbf {\bibinfo {volume}
  {84}},\ \bibinfo {pages} {205128} (\bibinfo {year} {2011})}\BibitemShut
  {NoStop}%
\bibitem [{\citenamefont {Bender}\ and\ \citenamefont
  {Boettcher}(1998)}]{PhysRevLett.80.5243}%
  \BibitemOpen
  \bibfield  {author} {\bibinfo {author} {\bibfnamefont {C.~M.}\ \bibnamefont
  {Bender}}\ and\ \bibinfo {author} {\bibfnamefont {S.}~\bibnamefont
  {Boettcher}},\ }\bibfield  {title} {\bibinfo {title} {Real spectra in
  non-hermitian hamiltonians having pt symmetry},\ }\href
  {https://doi.org/10.1103/PhysRevLett.80.5243} {\bibfield  {journal} {\bibinfo
   {journal} {Phys. Rev. Lett.}\ }\textbf {\bibinfo {volume} {80}},\ \bibinfo
  {pages} {5243} (\bibinfo {year} {1998})}\BibitemShut {NoStop}%
\bibitem [{\citenamefont {Denner}\ \emph {et~al.}(2021)\citenamefont {Denner},
  \citenamefont {Skurativska}, \citenamefont {Schindler}, \citenamefont
  {Fischer}, \citenamefont {Thomale}, \citenamefont {Bzdušek},\ and\
  \citenamefont {Neupert}}]{Denner2021}%
  \BibitemOpen
  \bibfield  {author} {\bibinfo {author} {\bibfnamefont {M.~M.}\ \bibnamefont
  {Denner}}, \bibinfo {author} {\bibfnamefont {A.}~\bibnamefont {Skurativska}},
  \bibinfo {author} {\bibfnamefont {F.}~\bibnamefont {Schindler}}, \bibinfo
  {author} {\bibfnamefont {M.~H.}\ \bibnamefont {Fischer}}, \bibinfo {author}
  {\bibfnamefont {R.}~\bibnamefont {Thomale}}, \bibinfo {author} {\bibfnamefont
  {T.}~\bibnamefont {Bzdušek}},\ and\ \bibinfo {author} {\bibfnamefont
  {T.}~\bibnamefont {Neupert}},\ }\bibfield  {title} {\bibinfo {title}
  {Exceptional topological insulators},\ }\bibfield  {journal} {\bibinfo
  {journal} {Nature Communications}\ }\textbf {\bibinfo {volume} {12}},\ \href
  {https://doi.org/10.1038/s41467-021-25947-z} {10.1038/s41467-021-25947-z}
  (\bibinfo {year} {2021})\BibitemShut {NoStop}%
\bibitem [{\citenamefont {Hyart}\ and\ \citenamefont
  {Lado}(2022)}]{PhysRevResearch.4.L012006}%
  \BibitemOpen
  \bibfield  {author} {\bibinfo {author} {\bibfnamefont {T.}~\bibnamefont
  {Hyart}}\ and\ \bibinfo {author} {\bibfnamefont {J.~L.}\ \bibnamefont
  {Lado}},\ }\bibfield  {title} {\bibinfo {title} {Non-hermitian many-body
  topological excitations in interacting quantum dots},\ }\href
  {https://doi.org/10.1103/PhysRevResearch.4.L012006} {\bibfield  {journal}
  {\bibinfo  {journal} {Phys. Rev. Res.}\ }\textbf {\bibinfo {volume} {4}},\
  \bibinfo {pages} {L012006} (\bibinfo {year} {2022})}\BibitemShut {NoStop}%
\bibitem [{\citenamefont {Takata}\ and\ \citenamefont {Notomi}(2018)}]{takata}%
  \BibitemOpen
  \bibfield  {author} {\bibinfo {author} {\bibfnamefont {K.}~\bibnamefont
  {Takata}}\ and\ \bibinfo {author} {\bibfnamefont {M.}~\bibnamefont
  {Notomi}},\ }\bibfield  {title} {\bibinfo {title} {Photonic topological
  insulating phase induced solely by gain and loss},\ }\href
  {https://doi.org/10.1103/PhysRevLett.121.213902} {\bibfield  {journal}
  {\bibinfo  {journal} {Phys. Rev. Lett.}\ }\textbf {\bibinfo {volume} {121}},\
  \bibinfo {pages} {213902} (\bibinfo {year} {2018})}\BibitemShut {NoStop}%
\bibitem [{\citenamefont {Ghatak}\ \emph {et~al.}(2020)\citenamefont {Ghatak},
  \citenamefont {Brandenbourger}, \citenamefont {van Wezel},\ and\
  \citenamefont {Coulais}}]{ghatak}%
  \BibitemOpen
  \bibfield  {author} {\bibinfo {author} {\bibfnamefont {A.}~\bibnamefont
  {Ghatak}}, \bibinfo {author} {\bibfnamefont {M.}~\bibnamefont
  {Brandenbourger}}, \bibinfo {author} {\bibfnamefont {J.}~\bibnamefont {van
  Wezel}},\ and\ \bibinfo {author} {\bibfnamefont {C.}~\bibnamefont
  {Coulais}},\ }\bibfield  {title} {\bibinfo {title} {Observation of
  non-hermitian topology and its bulk–edge correspondence in an active
  mechanical metamaterial},\ }\href {https://doi.org/10.1073/pnas.2010580117}
  {\bibfield  {journal} {\bibinfo  {journal} {Proceedings of the National
  Academy of Sciences}\ }\textbf {\bibinfo {volume} {117}},\ \bibinfo {pages}
  {29561–29568} (\bibinfo {year} {2020})}\BibitemShut {NoStop}%
\bibitem [{\citenamefont {Zhao}\ \emph {et~al.}(2019)\citenamefont {Zhao},
  \citenamefont {Qiao}, \citenamefont {Wu}, \citenamefont {Midya},
  \citenamefont {Longhi},\ and\ \citenamefont {Feng}}]{Zhao2019}%
  \BibitemOpen
  \bibfield  {author} {\bibinfo {author} {\bibfnamefont {H.}~\bibnamefont
  {Zhao}}, \bibinfo {author} {\bibfnamefont {X.}~\bibnamefont {Qiao}}, \bibinfo
  {author} {\bibfnamefont {T.}~\bibnamefont {Wu}}, \bibinfo {author}
  {\bibfnamefont {B.}~\bibnamefont {Midya}}, \bibinfo {author} {\bibfnamefont
  {S.}~\bibnamefont {Longhi}},\ and\ \bibinfo {author} {\bibfnamefont
  {L.}~\bibnamefont {Feng}},\ }\bibfield  {title} {\bibinfo {title}
  {Non-hermitian topological light steering},\ }\href
  {https://doi.org/10.1126/science.aay1064} {\bibfield  {journal} {\bibinfo
  {journal} {Science}\ }\textbf {\bibinfo {volume} {365}},\ \bibinfo {pages}
  {1163} (\bibinfo {year} {2019})}\BibitemShut {NoStop}%
\bibitem [{\citenamefont {Shalaev}\ \emph {et~al.}(2018)\citenamefont
  {Shalaev}, \citenamefont {Walasik}, \citenamefont {Tsukernik}, \citenamefont
  {Xu},\ and\ \citenamefont {Litchinitser}}]{Shalaev2018}%
  \BibitemOpen
  \bibfield  {author} {\bibinfo {author} {\bibfnamefont {M.~I.}\ \bibnamefont
  {Shalaev}}, \bibinfo {author} {\bibfnamefont {W.}~\bibnamefont {Walasik}},
  \bibinfo {author} {\bibfnamefont {A.}~\bibnamefont {Tsukernik}}, \bibinfo
  {author} {\bibfnamefont {Y.}~\bibnamefont {Xu}},\ and\ \bibinfo {author}
  {\bibfnamefont {N.~M.}\ \bibnamefont {Litchinitser}},\ }\bibfield  {title}
  {\bibinfo {title} {Robust topologically protected transport in photonic
  crystals at telecommunication wavelengths},\ }\href
  {https://doi.org/10.1038/s41565-018-0297-6} {\bibfield  {journal} {\bibinfo
  {journal} {Nature Nanotechnology}\ }\textbf {\bibinfo {volume} {14}},\
  \bibinfo {pages} {31} (\bibinfo {year} {2018})}\BibitemShut {NoStop}%
\bibitem [{\citenamefont {Arora}\ \emph {et~al.}(2021)\citenamefont {Arora},
  \citenamefont {Bauer}, \citenamefont {Barczyk}, \citenamefont {Verhagen},\
  and\ \citenamefont {Kuipers}}]{Arora2021}%
  \BibitemOpen
  \bibfield  {author} {\bibinfo {author} {\bibfnamefont {S.}~\bibnamefont
  {Arora}}, \bibinfo {author} {\bibfnamefont {T.}~\bibnamefont {Bauer}},
  \bibinfo {author} {\bibfnamefont {R.}~\bibnamefont {Barczyk}}, \bibinfo
  {author} {\bibfnamefont {E.}~\bibnamefont {Verhagen}},\ and\ \bibinfo
  {author} {\bibfnamefont {L.}~\bibnamefont {Kuipers}},\ }\bibfield  {title}
  {\bibinfo {title} {Direct quantification of topological protection in
  symmetry-protected photonic edge states at telecom wavelengths},\ }\bibfield
  {journal} {\bibinfo  {journal} {Light: Science and Applications}\ }\textbf
  {\bibinfo {volume} {10}},\ \href {https://doi.org/10.1038/s41377-020-00458-6}
  {10.1038/s41377-020-00458-6} (\bibinfo {year} {2021})\BibitemShut {NoStop}%
\bibitem [{\citenamefont {Bogaerts}\ \emph {et~al.}(2020)\citenamefont
  {Bogaerts}, \citenamefont {P{\'{e}}rez}, \citenamefont {Capmany},
  \citenamefont {Miller}, \citenamefont {Poon}, \citenamefont {Englund},
  \citenamefont {Morichetti},\ and\ \citenamefont {Melloni}}]{Bogaerts2020}%
  \BibitemOpen
  \bibfield  {author} {\bibinfo {author} {\bibfnamefont {W.}~\bibnamefont
  {Bogaerts}}, \bibinfo {author} {\bibfnamefont {D.}~\bibnamefont
  {P{\'{e}}rez}}, \bibinfo {author} {\bibfnamefont {J.}~\bibnamefont
  {Capmany}}, \bibinfo {author} {\bibfnamefont {D.~A.~B.}\ \bibnamefont
  {Miller}}, \bibinfo {author} {\bibfnamefont {J.}~\bibnamefont {Poon}},
  \bibinfo {author} {\bibfnamefont {D.}~\bibnamefont {Englund}}, \bibinfo
  {author} {\bibfnamefont {F.}~\bibnamefont {Morichetti}},\ and\ \bibinfo
  {author} {\bibfnamefont {A.}~\bibnamefont {Melloni}},\ }\bibfield  {title}
  {\bibinfo {title} {Programmable photonic circuits},\ }\href
  {https://doi.org/10.1038/s41586-020-2764-0} {\bibfield  {journal} {\bibinfo
  {journal} {Nature}\ }\textbf {\bibinfo {volume} {586}},\ \bibinfo {pages}
  {207} (\bibinfo {year} {2020})}\BibitemShut {NoStop}%
\bibitem [{\citenamefont {P{\'{e}}rez-L{\'{o}}pez}\ \emph
  {et~al.}(2020)\citenamefont {P{\'{e}}rez-L{\'{o}}pez}, \citenamefont
  {L{\'{o}}pez}, \citenamefont {DasMahapatra},\ and\ \citenamefont
  {Capmany}}]{PrezLpez2020}%
  \BibitemOpen
  \bibfield  {author} {\bibinfo {author} {\bibfnamefont {D.}~\bibnamefont
  {P{\'{e}}rez-L{\'{o}}pez}}, \bibinfo {author} {\bibfnamefont
  {A.}~\bibnamefont {L{\'{o}}pez}}, \bibinfo {author} {\bibfnamefont
  {P.}~\bibnamefont {DasMahapatra}},\ and\ \bibinfo {author} {\bibfnamefont
  {J.}~\bibnamefont {Capmany}},\ }\bibfield  {title} {\bibinfo {title}
  {Multipurpose self-configuration of programmable photonic circuits},\
  }\bibfield  {journal} {\bibinfo  {journal} {Nature Communications}\ }\textbf
  {\bibinfo {volume} {11}},\ \href {https://doi.org/10.1038/s41467-020-19608-w}
  {10.1038/s41467-020-19608-w} (\bibinfo {year} {2020})\BibitemShut {NoStop}%
\bibitem [{\citenamefont {Harris}\ \emph {et~al.}(2018)\citenamefont {Harris},
  \citenamefont {Carolan}, \citenamefont {Bunandar}, \citenamefont {Prabhu},
  \citenamefont {Hochberg}, \citenamefont {Baehr-Jones}, \citenamefont {Fanto},
  \citenamefont {Smith}, \citenamefont {Tison}, \citenamefont {Alsing},\ and\
  \citenamefont {Englund}}]{Harris2018}%
  \BibitemOpen
  \bibfield  {author} {\bibinfo {author} {\bibfnamefont {N.~C.}\ \bibnamefont
  {Harris}}, \bibinfo {author} {\bibfnamefont {J.}~\bibnamefont {Carolan}},
  \bibinfo {author} {\bibfnamefont {D.}~\bibnamefont {Bunandar}}, \bibinfo
  {author} {\bibfnamefont {M.}~\bibnamefont {Prabhu}}, \bibinfo {author}
  {\bibfnamefont {M.}~\bibnamefont {Hochberg}}, \bibinfo {author}
  {\bibfnamefont {T.}~\bibnamefont {Baehr-Jones}}, \bibinfo {author}
  {\bibfnamefont {M.~L.}\ \bibnamefont {Fanto}}, \bibinfo {author}
  {\bibfnamefont {A.~M.}\ \bibnamefont {Smith}}, \bibinfo {author}
  {\bibfnamefont {C.~C.}\ \bibnamefont {Tison}}, \bibinfo {author}
  {\bibfnamefont {P.~M.}\ \bibnamefont {Alsing}},\ and\ \bibinfo {author}
  {\bibfnamefont {D.}~\bibnamefont {Englund}},\ }\bibfield  {title} {\bibinfo
  {title} {Linear programmable nanophotonic processors},\ }\href
  {https://doi.org/10.1364/optica.5.001623} {\bibfield  {journal} {\bibinfo
  {journal} {Optica}\ }\textbf {\bibinfo {volume} {5}},\ \bibinfo {pages}
  {1623} (\bibinfo {year} {2018})}\BibitemShut {NoStop}%
\bibitem [{\citenamefont {Clements}\ \emph {et~al.}(2016)\citenamefont
  {Clements}, \citenamefont {Humphreys}, \citenamefont {Metcalf}, \citenamefont
  {Kolthammer},\ and\ \citenamefont {Walsmley}}]{Clements2016}%
  \BibitemOpen
  \bibfield  {author} {\bibinfo {author} {\bibfnamefont {W.~R.}\ \bibnamefont
  {Clements}}, \bibinfo {author} {\bibfnamefont {P.~C.}\ \bibnamefont
  {Humphreys}}, \bibinfo {author} {\bibfnamefont {B.~J.}\ \bibnamefont
  {Metcalf}}, \bibinfo {author} {\bibfnamefont {W.~S.}\ \bibnamefont
  {Kolthammer}},\ and\ \bibinfo {author} {\bibfnamefont {I.~A.}\ \bibnamefont
  {Walsmley}},\ }\bibfield  {title} {\bibinfo {title} {Optimal design for
  universal multiport interferometers},\ }\href
  {https://doi.org/10.1364/optica.3.001460} {\bibfield  {journal} {\bibinfo
  {journal} {Optica}\ }\textbf {\bibinfo {volume} {3}},\ \bibinfo {pages}
  {1460} (\bibinfo {year} {2016})}\BibitemShut {NoStop}%
\bibitem [{\citenamefont {On}\ \emph {et~al.}(2024)\citenamefont {On},
  \citenamefont {Ashtiani}, \citenamefont {Sanchez-Jacome}, \citenamefont
  {Perez-Lopez}, \citenamefont {Yoo},\ and\ \citenamefont
  {Blanco-Redondo}}]{andreas}%
  \BibitemOpen
  \bibfield  {author} {\bibinfo {author} {\bibfnamefont {M.~B.}\ \bibnamefont
  {On}}, \bibinfo {author} {\bibfnamefont {F.}~\bibnamefont {Ashtiani}},
  \bibinfo {author} {\bibfnamefont {D.}~\bibnamefont {Sanchez-Jacome}},
  \bibinfo {author} {\bibfnamefont {D.}~\bibnamefont {Perez-Lopez}}, \bibinfo
  {author} {\bibfnamefont {S.~J.~B.}\ \bibnamefont {Yoo}},\ and\ \bibinfo
  {author} {\bibfnamefont {A.}~\bibnamefont {Blanco-Redondo}},\ }\bibfield
  {title} {\bibinfo {title} {Programmable integrated photonics for topological
  hamiltonians},\ }\bibfield  {journal} {\bibinfo  {journal} {Nature
  Communications}\ }\textbf {\bibinfo {volume} {15}},\ \href
  {https://doi.org/10.1038/s41467-024-44939-3} {10.1038/s41467-024-44939-3}
  (\bibinfo {year} {2024})\BibitemShut {NoStop}%
\bibitem [{\citenamefont {Liu}\ \emph {et~al.}(2021{\natexlab{a}})\citenamefont
  {Liu}, \citenamefont {Zhou},\ and\ \citenamefont
  {Chen}}]{PhysRevB.104.024201}%
  \BibitemOpen
  \bibfield  {author} {\bibinfo {author} {\bibfnamefont {Y.}~\bibnamefont
  {Liu}}, \bibinfo {author} {\bibfnamefont {Q.}~\bibnamefont {Zhou}},\ and\
  \bibinfo {author} {\bibfnamefont {S.}~\bibnamefont {Chen}},\ }\bibfield
  {title} {\bibinfo {title} {Localization transition, spectrum structure, and
  winding numbers for one-dimensional non-hermitian quasicrystals},\ }\href
  {https://doi.org/10.1103/PhysRevB.104.024201} {\bibfield  {journal} {\bibinfo
   {journal} {Phys. Rev. B}\ }\textbf {\bibinfo {volume} {104}},\ \bibinfo
  {pages} {024201} (\bibinfo {year} {2021}{\natexlab{a}})}\BibitemShut
  {NoStop}%
\bibitem [{\citenamefont {Liu}\ \emph {et~al.}(2021{\natexlab{b}})\citenamefont
  {Liu}, \citenamefont {Wang}, \citenamefont {Liu}, \citenamefont {Zhou},\ and\
  \citenamefont {Chen}}]{PhysRevB.103.014203}%
  \BibitemOpen
  \bibfield  {author} {\bibinfo {author} {\bibfnamefont {Y.}~\bibnamefont
  {Liu}}, \bibinfo {author} {\bibfnamefont {Y.}~\bibnamefont {Wang}}, \bibinfo
  {author} {\bibfnamefont {X.-J.}\ \bibnamefont {Liu}}, \bibinfo {author}
  {\bibfnamefont {Q.}~\bibnamefont {Zhou}},\ and\ \bibinfo {author}
  {\bibfnamefont {S.}~\bibnamefont {Chen}},\ }\bibfield  {title} {\bibinfo
  {title} {Exact mobility edges, $\mathcal{PT}$-symmetry breaking, and skin
  effect in one-dimensional non-hermitian quasicrystals},\ }\href
  {https://doi.org/10.1103/PhysRevB.103.014203} {\bibfield  {journal} {\bibinfo
   {journal} {Phys. Rev. B}\ }\textbf {\bibinfo {volume} {103}},\ \bibinfo
  {pages} {014203} (\bibinfo {year} {2021}{\natexlab{b}})}\BibitemShut
  {NoStop}%
\bibitem [{\citenamefont {Longhi}(2019)}]{PhysRevB.100.125157}%
  \BibitemOpen
  \bibfield  {author} {\bibinfo {author} {\bibfnamefont {S.}~\bibnamefont
  {Longhi}},\ }\bibfield  {title} {\bibinfo {title} {Metal-insulator phase
  transition in a non-hermitian aubry-andr\'e-harper model},\ }\href
  {https://doi.org/10.1103/PhysRevB.100.125157} {\bibfield  {journal} {\bibinfo
   {journal} {Phys. Rev. B}\ }\textbf {\bibinfo {volume} {100}},\ \bibinfo
  {pages} {125157} (\bibinfo {year} {2019})}\BibitemShut {NoStop}%
\bibitem [{\citenamefont {Brzezicki}\ and\ \citenamefont
  {Hyart}(2019)}]{PhysRevB.100.161105}%
  \BibitemOpen
  \bibfield  {author} {\bibinfo {author} {\bibfnamefont {W.}~\bibnamefont
  {Brzezicki}}\ and\ \bibinfo {author} {\bibfnamefont {T.}~\bibnamefont
  {Hyart}},\ }\bibfield  {title} {\bibinfo {title} {Hidden chern number in
  one-dimensional non-hermitian chiral-symmetric systems},\ }\href
  {https://doi.org/10.1103/PhysRevB.100.161105} {\bibfield  {journal} {\bibinfo
   {journal} {Phys. Rev. B}\ }\textbf {\bibinfo {volume} {100}},\ \bibinfo
  {pages} {161105} (\bibinfo {year} {2019})}\BibitemShut {NoStop}%
\bibitem [{\citenamefont {Gao}\ \emph {et~al.}(2020)\citenamefont {Gao},
  \citenamefont {Xue}, \citenamefont {Wang}, \citenamefont {Gu}, \citenamefont
  {Liu}, \citenamefont {Zhu},\ and\ \citenamefont
  {Zhang}}]{PhysRevB.101.180303}%
  \BibitemOpen
  \bibfield  {author} {\bibinfo {author} {\bibfnamefont {H.}~\bibnamefont
  {Gao}}, \bibinfo {author} {\bibfnamefont {H.}~\bibnamefont {Xue}}, \bibinfo
  {author} {\bibfnamefont {Q.}~\bibnamefont {Wang}}, \bibinfo {author}
  {\bibfnamefont {Z.}~\bibnamefont {Gu}}, \bibinfo {author} {\bibfnamefont
  {T.}~\bibnamefont {Liu}}, \bibinfo {author} {\bibfnamefont {J.}~\bibnamefont
  {Zhu}},\ and\ \bibinfo {author} {\bibfnamefont {B.}~\bibnamefont {Zhang}},\
  }\bibfield  {title} {\bibinfo {title} {Observation of topological edge states
  induced solely by non-hermiticity in an acoustic crystal},\ }\href
  {https://doi.org/10.1103/PhysRevB.101.180303} {\bibfield  {journal} {\bibinfo
   {journal} {Phys. Rev. B}\ }\textbf {\bibinfo {volume} {101}},\ \bibinfo
  {pages} {180303} (\bibinfo {year} {2020})}\BibitemShut {NoStop}%
\bibitem [{\citenamefont {Liu}\ \emph {et~al.}(2020)\citenamefont {Liu},
  \citenamefont {Ma}, \citenamefont {Yang}, \citenamefont {Zhang},
  \citenamefont {Gao}, \citenamefont {Xiang}, \citenamefont {Cui},\ and\
  \citenamefont {Zhang}}]{PhysRevApplied.13.014047}%
  \BibitemOpen
  \bibfield  {author} {\bibinfo {author} {\bibfnamefont {S.}~\bibnamefont
  {Liu}}, \bibinfo {author} {\bibfnamefont {S.}~\bibnamefont {Ma}}, \bibinfo
  {author} {\bibfnamefont {C.}~\bibnamefont {Yang}}, \bibinfo {author}
  {\bibfnamefont {L.}~\bibnamefont {Zhang}}, \bibinfo {author} {\bibfnamefont
  {W.}~\bibnamefont {Gao}}, \bibinfo {author} {\bibfnamefont {Y.~J.}\
  \bibnamefont {Xiang}}, \bibinfo {author} {\bibfnamefont {T.~J.}\ \bibnamefont
  {Cui}},\ and\ \bibinfo {author} {\bibfnamefont {S.}~\bibnamefont {Zhang}},\
  }\bibfield  {title} {\bibinfo {title} {Gain- and loss-induced topological
  insulating phase in a non-hermitian electrical circuit},\ }\href
  {https://doi.org/10.1103/PhysRevApplied.13.014047} {\bibfield  {journal}
  {\bibinfo  {journal} {Phys. Rev. Appl.}\ }\textbf {\bibinfo {volume} {13}},\
  \bibinfo {pages} {014047} (\bibinfo {year} {2020})}\BibitemShut {NoStop}%
\bibitem [{\citenamefont {Chen}\ \emph {et~al.}(2023)\citenamefont {Chen},
  \citenamefont {Song},\ and\ \citenamefont {Lado}}]{PhysRevLett.130.100401}%
  \BibitemOpen
  \bibfield  {author} {\bibinfo {author} {\bibfnamefont {G.}~\bibnamefont
  {Chen}}, \bibinfo {author} {\bibfnamefont {F.}~\bibnamefont {Song}},\ and\
  \bibinfo {author} {\bibfnamefont {J.~L.}\ \bibnamefont {Lado}},\ }\bibfield
  {title} {\bibinfo {title} {Topological spin excitations in non-hermitian spin
  chains with a generalized kernel polynomial algorithm},\ }\href
  {https://doi.org/10.1103/PhysRevLett.130.100401} {\bibfield  {journal}
  {\bibinfo  {journal} {Phys. Rev. Lett.}\ }\textbf {\bibinfo {volume} {130}},\
  \bibinfo {pages} {100401} (\bibinfo {year} {2023})}\BibitemShut {NoStop}%
\bibitem [{\citenamefont {Zhu}\ \emph {et~al.}(2023)\citenamefont {Zhu},
  \citenamefont {Lang}, \citenamefont {Wang}, \citenamefont {Wang},\ and\
  \citenamefont {Chong}}]{zhulang}%
  \BibitemOpen
  \bibfield  {author} {\bibinfo {author} {\bibfnamefont {B.}~\bibnamefont
  {Zhu}}, \bibinfo {author} {\bibfnamefont {L.-J.}\ \bibnamefont {Lang}},
  \bibinfo {author} {\bibfnamefont {Q.}~\bibnamefont {Wang}}, \bibinfo {author}
  {\bibfnamefont {Q.~J.}\ \bibnamefont {Wang}},\ and\ \bibinfo {author}
  {\bibfnamefont {Y.~D.}\ \bibnamefont {Chong}},\ }\bibfield  {title} {\bibinfo
  {title} {Topological transitions with an imaginary aubry-andr\'e-harper
  potential},\ }\href {https://doi.org/10.1103/PhysRevResearch.5.023044}
  {\bibfield  {journal} {\bibinfo  {journal} {Phys. Rev. Res.}\ }\textbf
  {\bibinfo {volume} {5}},\ \bibinfo {pages} {023044} (\bibinfo {year}
  {2023})}\BibitemShut {NoStop}%
\bibitem [{\citenamefont {Ganeshan}\ \emph {et~al.}(2013)\citenamefont
  {Ganeshan}, \citenamefont {Sun},\ and\ \citenamefont
  {Das~Sarma}}]{Ganeshan2013}%
  \BibitemOpen
  \bibfield  {author} {\bibinfo {author} {\bibfnamefont {S.}~\bibnamefont
  {Ganeshan}}, \bibinfo {author} {\bibfnamefont {K.}~\bibnamefont {Sun}},\ and\
  \bibinfo {author} {\bibfnamefont {S.}~\bibnamefont {Das~Sarma}},\ }\bibfield
  {title} {\bibinfo {title} {Topological zero-energy modes in gapless
  commensurate aubry-andr\'e-harper models},\ }\href
  {https://doi.org/10.1103/PhysRevLett.110.180403} {\bibfield  {journal}
  {\bibinfo  {journal} {Phys. Rev. Lett.}\ }\textbf {\bibinfo {volume} {110}},\
  \bibinfo {pages} {180403} (\bibinfo {year} {2013})}\BibitemShut {NoStop}%
\bibitem [{\citenamefont {Jitomirskaya}(1999)}]{Jitomirskaya1999}%
  \BibitemOpen
  \bibfield  {author} {\bibinfo {author} {\bibfnamefont {S.~Y.}\ \bibnamefont
  {Jitomirskaya}},\ }\bibfield  {title} {\bibinfo {title} {Metal-insulator
  transition for the almost mathieu operator},\ }\href
  {https://doi.org/10.2307/121066} {\bibfield  {journal} {\bibinfo  {journal}
  {The Annals of Mathematics}\ }\textbf {\bibinfo {volume} {150}},\ \bibinfo
  {pages} {1159} (\bibinfo {year} {1999})}\BibitemShut {NoStop}%
\bibitem [{\citenamefont {Wang}\ \emph {et~al.}(2020)\citenamefont {Wang},
  \citenamefont {Xia}, \citenamefont {Zhang}, \citenamefont {Yao},
  \citenamefont {Chen}, \citenamefont {You}, \citenamefont {Zhou},\ and\
  \citenamefont {Liu}}]{PhysRevLett.125.196604}%
  \BibitemOpen
  \bibfield  {author} {\bibinfo {author} {\bibfnamefont {Y.}~\bibnamefont
  {Wang}}, \bibinfo {author} {\bibfnamefont {X.}~\bibnamefont {Xia}}, \bibinfo
  {author} {\bibfnamefont {L.}~\bibnamefont {Zhang}}, \bibinfo {author}
  {\bibfnamefont {H.}~\bibnamefont {Yao}}, \bibinfo {author} {\bibfnamefont
  {S.}~\bibnamefont {Chen}}, \bibinfo {author} {\bibfnamefont {J.}~\bibnamefont
  {You}}, \bibinfo {author} {\bibfnamefont {Q.}~\bibnamefont {Zhou}},\ and\
  \bibinfo {author} {\bibfnamefont {X.-J.}\ \bibnamefont {Liu}},\ }\bibfield
  {title} {\bibinfo {title} {One-dimensional quasiperiodic mosaic lattice with
  exact mobility edges},\ }\href
  {https://doi.org/10.1103/PhysRevLett.125.196604} {\bibfield  {journal}
  {\bibinfo  {journal} {Phys. Rev. Lett.}\ }\textbf {\bibinfo {volume} {125}},\
  \bibinfo {pages} {196604} (\bibinfo {year} {2020})}\BibitemShut {NoStop}%
\bibitem [{\citenamefont {Kohmoto}\ \emph {et~al.}(1983)\citenamefont
  {Kohmoto}, \citenamefont {Kadanoff},\ and\ \citenamefont
  {Tang}}]{PhysRevLett.50.1870}%
  \BibitemOpen
  \bibfield  {author} {\bibinfo {author} {\bibfnamefont {M.}~\bibnamefont
  {Kohmoto}}, \bibinfo {author} {\bibfnamefont {L.~P.}\ \bibnamefont
  {Kadanoff}},\ and\ \bibinfo {author} {\bibfnamefont {C.}~\bibnamefont
  {Tang}},\ }\bibfield  {title} {\bibinfo {title} {Localization problem in one
  dimension: Mapping and escape},\ }\href
  {https://doi.org/10.1103/PhysRevLett.50.1870} {\bibfield  {journal} {\bibinfo
   {journal} {Phys. Rev. Lett.}\ }\textbf {\bibinfo {volume} {50}},\ \bibinfo
  {pages} {1870} (\bibinfo {year} {1983})}\BibitemShut {NoStop}%
\bibitem [{\citenamefont {Khosravian}\ and\ \citenamefont
  {Lado}(2021)}]{PhysRevResearch.3.013262}%
  \BibitemOpen
  \bibfield  {author} {\bibinfo {author} {\bibfnamefont {M.}~\bibnamefont
  {Khosravian}}\ and\ \bibinfo {author} {\bibfnamefont {J.~L.}\ \bibnamefont
  {Lado}},\ }\bibfield  {title} {\bibinfo {title} {Quasiperiodic criticality
  and spin-triplet superconductivity in superconductor-antiferromagnet moir\'e
  patterns},\ }\href {https://doi.org/10.1103/PhysRevResearch.3.013262}
  {\bibfield  {journal} {\bibinfo  {journal} {Phys. Rev. Res.}\ }\textbf
  {\bibinfo {volume} {3}},\ \bibinfo {pages} {013262} (\bibinfo {year}
  {2021})}\BibitemShut {NoStop}%
\bibitem [{\citenamefont {Ostlund}\ \emph {et~al.}(1983)\citenamefont
  {Ostlund}, \citenamefont {Pandit}, \citenamefont {Rand}, \citenamefont
  {Schellnhuber},\ and\ \citenamefont {Siggia}}]{PhysRevLett.50.1873}%
  \BibitemOpen
  \bibfield  {author} {\bibinfo {author} {\bibfnamefont {S.}~\bibnamefont
  {Ostlund}}, \bibinfo {author} {\bibfnamefont {R.}~\bibnamefont {Pandit}},
  \bibinfo {author} {\bibfnamefont {D.}~\bibnamefont {Rand}}, \bibinfo {author}
  {\bibfnamefont {H.~J.}\ \bibnamefont {Schellnhuber}},\ and\ \bibinfo {author}
  {\bibfnamefont {E.~D.}\ \bibnamefont {Siggia}},\ }\bibfield  {title}
  {\bibinfo {title} {One-dimensional schr\"odinger equation with an almost
  periodic potential},\ }\href {https://doi.org/10.1103/PhysRevLett.50.1873}
  {\bibfield  {journal} {\bibinfo  {journal} {Phys. Rev. Lett.}\ }\textbf
  {\bibinfo {volume} {50}},\ \bibinfo {pages} {1873} (\bibinfo {year}
  {1983})}\BibitemShut {NoStop}%
\bibitem [{\citenamefont {Anderson}(1958)}]{PhysRev.109.1492}%
  \BibitemOpen
  \bibfield  {author} {\bibinfo {author} {\bibfnamefont {P.~W.}\ \bibnamefont
  {Anderson}},\ }\bibfield  {title} {\bibinfo {title} {Absence of diffusion in
  certain random lattices},\ }\href {https://doi.org/10.1103/PhysRev.109.1492}
  {\bibfield  {journal} {\bibinfo  {journal} {Phys. Rev.}\ }\textbf {\bibinfo
  {volume} {109}},\ \bibinfo {pages} {1492} (\bibinfo {year}
  {1958})}\BibitemShut {NoStop}%
\bibitem [{\citenamefont {Lee}\ and\ \citenamefont
  {Ramakrishnan}(1985)}]{RevModPhys.57.287}%
  \BibitemOpen
  \bibfield  {author} {\bibinfo {author} {\bibfnamefont {P.~A.}\ \bibnamefont
  {Lee}}\ and\ \bibinfo {author} {\bibfnamefont {T.~V.}\ \bibnamefont
  {Ramakrishnan}},\ }\bibfield  {title} {\bibinfo {title} {Disordered
  electronic systems},\ }\href {https://doi.org/10.1103/RevModPhys.57.287}
  {\bibfield  {journal} {\bibinfo  {journal} {Rev. Mod. Phys.}\ }\textbf
  {\bibinfo {volume} {57}},\ \bibinfo {pages} {287} (\bibinfo {year}
  {1985})}\BibitemShut {NoStop}%
\bibitem [{\citenamefont {Biddle}\ and\ \citenamefont
  {Das~Sarma}(2010)}]{Biddle2010}%
  \BibitemOpen
  \bibfield  {author} {\bibinfo {author} {\bibfnamefont {J.}~\bibnamefont
  {Biddle}}\ and\ \bibinfo {author} {\bibfnamefont {S.}~\bibnamefont
  {Das~Sarma}},\ }\bibfield  {title} {\bibinfo {title} {Predicted mobility
  edges in one-dimensional incommensurate optical lattices: An exactly solvable
  model of anderson localization},\ }\href
  {https://doi.org/10.1103/PhysRevLett.104.070601} {\bibfield  {journal}
  {\bibinfo  {journal} {Phys. Rev. Lett.}\ }\textbf {\bibinfo {volume} {104}},\
  \bibinfo {pages} {070601} (\bibinfo {year} {2010})}\BibitemShut {NoStop}%
\end{thebibliography}%

 \end{document}